\documentclass[12pt,preprint]{aastex}
\usepackage[yyyymmdd,hhmmss]{datetime}
\usepackage[margin=1.0in]{geometry}

\shorttitle{WRs in the Magellanic Clouds}

\slugcomment{Accepted by the Astrophysical Journal}
\shortauthors{Massey et al.}

\begin{document}

\title{A Modern Search for Wolf-Rayet Stars in the Magellanic Clouds: First Results\altaffilmark{1}}

\author{Philip Massey\altaffilmark{2}, Kathryn F. Neugent\altaffilmark{2}, Nidia Morrell\altaffilmark{3}, and D. John Hillier\altaffilmark{4}}

\altaffiltext{1}{This paper includes data gathered with the 1 m Swope and 6.5 m Magellan Telescopes located at Las Campanas Observatory, Chile.}
\altaffiltext{2}{Lowell Observatory, 1400 W Mars Hill Road, Flagstaff, AZ 86001; phil.massey@lowell.edu; \\kneugent@lowell.edu.}
\altaffiltext{3}{Las Campanas Observatory, Carnegie Observatories, Casilla 601, La Serena, Chile; nmorrell@lco.cl.}
\altaffiltext{4}{Department of Physics and Astronomy \& Pittsburgh Particle Physics, Astrophysics, and Cosmology Center (PITT PACC), University of Pittsburgh, Pittsburgh, PA 15260; hillier@pitt.edu.}

\begin{abstract}

Over the years, directed surveys and incidental spectroscopy have identified 12 Wolf-Rayet (WR) stars in the SMC and 139 in the LMC, numbers which are often described as ``{\it essentially} complete."  Yet, new WRs are discovered in the LMC almost yearly.  We have therefore initiated a new survey of both Magellanic Clouds using the same interference-filter imaging technique previously applied to M31 and M33.  We report on our first observing season, in which we have successfully
surveyed $\sim15$\% of our intended area of the SMC and LMC.  Spectroscopy has confirmed 9 newly found WRs in the LMC (a 6\% increase), including one of WO-type, only the third known in that galaxy and the second to be discovered recently. The other eight are WN3 stars that include an absorption component.  In two, the absorption is likely from  an O-type companion, but the other six are quite unusual.  Five would be classified naively as ``WN3+O3~V," but such a pairing is unlikely given the rarity of O3 stars, the short duration of this phase (which is incommensurate with the evolution of a companion to a WN star), and because these stars are considerably fainter than O3~V stars.  The sixth star may also fall into this category.  
CMFGEN modeling suggests these stars are hot, bolometrically luminous,  and N-rich like other WN3 stars, but lack the strong winds that characterize WNs.  Finally, we discuss two rare Of?p stars and four Of supergiants
we found, and propose that the B[e] star HD~38489 may have a WN companion.

\vskip -10pt
\end{abstract}


\section{Introduction}
\label{Sec-intro}

\vskip -10pt

Wolf-Rayet (WR) stars are the evolved descendants of massive O-type stars.  Mass loss during the main-sequence phase, possibly aided by episodic mass ejection during the Luminous Blue Variable (LBV) stage and/or Roche-lobe overflow in close binary systems, strips off the star's H-rich outer layers.  This mass loss, plus mixing from the interior, helps to reveal enhanced He and N (the products of CNO H-burning) at the surface. 
Such a star is identified spectroscopically as a WN-type WR.  If the star has sufficiently high mass, then additional evolution, mass loss, and mixing will lead to a WC-type WR, with enhanced C and O (the products of He-burning).  Further evolution may lead to one of the very rare WO-type WRs. The spectra of WRs are characterized by broad, strong emission lines as these lines are formed in an extended, expanding atmosphere/stellar wind; if absorption is present in the spectrum, it is usually
(but not always) due to a close OB companion.  Reviews are provided by Maeder \& Conti (1994), Crowther (2008) and Massey (2013), among others.

The relative number of WN- and WC-type WRs as a function of metallicity has long been used as a key diagnostic of massive star evolutionary models.  Main-sequence mass-loss rates are larger at
higher metallicities, as they are driven by radiation pressure acting on
highly ionized metal ions. (The metallicity-dependence of wind-driven mass loss was first offered as an explanation for the changing WC to WN number ratio by Vanbeveren \& Conti 1980.)    The conventional wisdom has long been that while single-star evolutionary models do a good job of matching the WC/WN ratio at lower metallicities (such as those found in the Magellanic Clouds), they fail at the higher metallicities characteristic of the center of M33, which has a metallicity that is approximately solar,  and M31, which has a metallicity that is approximately  $2\times$ solar.  Examples of this are shown by Massey \& Johnson (1998), Meynet \& Maeder (2005), and most recently by Neugent et al.\ (2012a). 

The linchpins for such comparisons at lower metallicities are the Magellanic Clouds.  They are the nearest star-forming galaxies to our own, and studies over the years have identified 139 WRs in the LMC (134 stars listed in the Breysacher et al.\ 1999 catalog [BAT99] plus 7 WRs subsequently discovered by various studies, minus 2 that have been demoted to Of-type; see Table 3 of Neugent et al.\ 2012b and references therein\footnote{Note that the reference for the discovery of [M2002] LMC 15666 as a WR star is incorrectly given in that table.  Instead,
the discovery should be credited to Gvaramadze et al.\ (2012), who reported the discovery of a WR star in the LMC, but did not provide any coordinates or cross-IDs in that brief conference proceeding.  Brian Skiff identified the object from their images, and this is the source of the information in Neugent et al.\ (2012b) and in SIMBAD.  It is the NE component of a 2\arcsec\ pair, with the companion a B0~V star.})  and 12 in the SMC (8 listed by Azzopardi \& Breysacher 1979a, plus 4 WRs subsequently discovered; see Table 1 of Massey et al.\ 2003 and references therein).  For years it has been commonly accepted that these numbers are {\it essentially} complete.   For instance, in their report of discovering two WRs in the LMC, Howarth \& Walborn (2012) suggested that perhaps as many as a dozen or so weak-lined WNEs (10\% of the LMC's total WR population) remained to be found, but no more.   
However, even before the Howarth \& Walborn (2012) paper appeared in print, Neugent et al.\ (2012b) announced the discovery of a very strong-lined WO-type WR in the LMC.  Similarly, Massey \& Duffy (2001) concluded that the completeness of their survey for WRs in the SMC 
could not ``preclude a WR star (or two) [from] having
been overlooked," a statement that proved prescient, as another SMC WN star was chanced upon within a year (Massey et al.\ 2003). 

These discoveries were unsettling, and forced us to examine how we came to know the WR content of the Magellanic Clouds, and what would be involved in conducting a more thorough survey, particularly in the LMC where the number of WR stars is large enough to provide robust statistics.  Some of the discoveries of WRs in the Clouds came about as part of general spectroscopy, while others came about as a result of directed objective prism searches. Of the 158 ``brightest stars" in the Magellanic Clouds, 15 were classified
as WR type (Feast et al.\ 1960) through various spectroscopic surveys.  An objective prism survey of the LMC aimed at finding WRs by Westerlund \& Rodgers (1959) resulted in the identification of 50 WRs, 30 of which were in common with those known from the HD catalog.   (See also Westerlund \& Smith 1964.)  Deeper and more complete surveys for WRs in the Magellanic Clouds were carried out by Azzopardi \& Breysacher (1979a, 1979b, 1980).  Their surveys employed an objective prism in combination with an interference filter that isolated the region around C~III $\lambda 4650$ and He~II $\lambda 4686$ (the two strongest optical emission lines in the spectra of WC- and WN-type WRs, respectively) in order to reduce problems with crowding and sky background that would have occurred with the use of the objective prism by itself.   
These studies added 4 additional WRs to the 4 that were previously known
in the SMC (Breysacher \& Westerlund 1978), 
and 17 additional WRs to the 80 known  in the LMC (Fehrenbach et al.\ 1976).  
It is worth noting that {\it all} of the new WRs found by Azzopardi \& Breysacher (1979a, 1979b, 1980) were of WN type.

Indeed, the difficulty in identifying unbiased samples of WRs has been described by Massey \& Johnson (1998): the strongest optical line in WC stars (typically C III $\lambda4650$) is about 4$\times$ stronger (on average) than that found in WN stars (He~II $\lambda$4686).  The weakest-lined WN stars have He~II $\lambda 4686$ equivalent widths of just $-$10~\AA, in contrast to the $-$50~\AA\ equivalent widths found in the weakest-lined WCs (see, e.g., Fig.~1 of Massey \& Johnson 1998).  Thus, a survey for WRs has to be sufficiently sensitive to detect weak-lined WNs if it is going to be useful for comparing with the predictions of the evolutionary models.   Armandroff \& Massey (1985) described a set of interference filters that has proven very effective at this task: three 50~\AA\ wide filters centered on C III $\lambda 4650$, He~II $\lambda 4686$, and neighboring continuum at $\lambda 4750$ are used to image a region with CCDs, and the brightness of objects
compared.  This was done  by Armandroff \& Massey (1985), Massey et al.\ (1986, 1992), and Massey \& Johnson (1998) to survey small regions of Local Group galaxies beyond the Magellanic Clouds (e.g., parts of  M31, M33, NGC~6822, IC~1613, IC 10, and NGC~6822) using the relatively tiny CCDs that were then available. Crowded-field photometry algorithms  (i.e., PSF-fitting with {\it DAOPHOT}, Stetson 1987) were then used to find WR candidates that were significantly brighter in one of the on-band filters compared to the 
expected photometric errors.  More
recently, we have been able to take advantage of CCD cameras with much larger fields of view to survey all of M33
(Neugent \& Massey 2011) and M31 (Neugent et al.\ 2012a).  These two surveys used  image-subtraction techniques to search for candidates in order to avoid the many false positives that plagued the photometry method.  Spectroscopic confirmation of these candidates demonstrated that we were finding WRs as weak-lined as any known, and indeed finding new Of-type stars with even smaller emission-lines fluxes, lending  some confidence that these surveys were sufficiently sensitive and deep to be detecting the vast majority of the WNs.

For the SMC, Massey \& Duffy (2001) undertook such a survey using a wide-area CCD on the CTIO Curtis Schmidt.  It 
covered 9.6 deg$^2$ and spectroscopy confirmed two new WR stars (both WNs),  bringing the total number of known WRs in
the SMC to 11, the result of photometry of over 1.6 million stars.    Still, the survey had some deficiencies: the pixel size with the instrument was 2\farcs3, reducing the precision in crowded regions.  The areal coverage, while large, did not cover all of the star-forming regions of the SMC.  As mentioned above, the next year saw the discovery of a 12th SMC WR (another WN) which had been overlooked in the Massey \& Duffy (2001) survey because of crowding.

For the LMC, in the 20 years between the Azzopardi \& Breysacher (1979b, 1980) survey and the compilation of the BAT99  ``Fourth Catalogue,"   the number of WRs known grew from 97 to 134 (roughly 40\%), mostly as a result of  accidental discovery through spectroscopy of stars in selected regions of the Clouds, with only 6 found as 
the result of new objective prism surveys for WRs (Morgan \& Good 1985, 1990).   Since BAT99, the number of known LMC WRs has grown to 
139 (an increase of  4\%), including two demotions of WRs to Of-type.
To us, our discovery (Neugent et al.\ 2012b) of a very rare (and very strong-lined) WO-type WR in Lucke-Hodge 41, a well-studied LMC OB association (harboring, among other things, another WR star and two LBVs, including the archetype S Doradus itself), seemed a wakeup call.  We are in the somewhat embarrassing position of knowing more about the WR content of
M33 and M31 (at distances of $\sim$ 800 kpc) than we do about our next-door neighbors, the Magellanic Clouds (at distances
of 50-60 kpc, i.e., $\sim 15 \times$ closer).  

The question, then, was what to do about it.  We decided to survey the Magellanic Clouds for WRs using the same method that we had so successfully employed in M31 and M33, using interference-filter imaging and image-subtraction
to identify WR candidates and then using spectroscopy to confirm and classify them.  The task, however, is quite daunting, as 
one of the things that makes the Magellanic Clouds so attractive---their closeness---also results in their very large angular sizes.  
The need, however,  is timely:  Improved evolutionary models have become
available from the Geneva group that actually now predict a significantly smaller WC/WN ratio ($\sim 0.1$) for the LMC than the ``observed" ratio (0.23).   Is the problem with the models, or is the problem that too many WNs have been missed in
past studies?  At the same time, the Cambridge STARS evolutionary models continue to improve and become increasingly available (see, e.g., Eldridge et al.\ 2008), allowing comparison with models that include the effects of close binary evolution as well.  And, of course, one never knows what surprises await in such new surveys, as we shall see.

Here we report the results of our first observing season.  Although we have only surveyed $\sim15$\% of the SMC and LMC, we have already discovered nine new WRs (all in the LMC), two interesting ``Of?p" stars,  four  Of-type supergiants and a new O4~V star.  Of the 9 new
WRs, the majority show strong early-type absorption.  Are these extremely massive binaries, or are they members of a newly discovered class of massive stars?   We describe the details of the new survey and follow-up spectroscopy
in Section~\ref{Sec-survey}, provide our new spectral types in Section~\ref{Sec-results},  demonstrate the sensitivity of our survey in Section~\ref{Sec-completeness}, and discuss the nature of our new discoveries in Section~\ref{Sec-sum}.

\section{The New Survey Begins}
\label{Sec-survey}
In designing our survey we were guided by wide-field (``parking-lot camera") images  of the SMC and LMC  described by Bothun \& Thompson (1988) and kindly made available by G. Bothun.  
We decided to survey a region extending 3\fdg0 in radius (28.3 deg$^2$) centered on 
$\alpha_{\rm 2000}=1^h08^m00^s, \delta_{\rm 2000}=-73\degr10\arcmin00\arcsec$ for the SMC, and a region extending 3\fdg5 in radius (38.5 deg$^2$) centered on $\alpha_{\rm 2000}=5^h18^m00^s, \delta_{\rm 2000}=-68\degr45\arcmin00\arcsec$  for the LMC.  These areas are shown in Figs.~\ref{fig:SMC} and
~\ref{fig:LMC}, respectively.  These regions encompass all of the HII regions shown in the parking-lot camera
H$\alpha$ images, as well as cataloged OB associations shown by Hodge (1985) and Lucke \& Hodge (1970).
Note that the area to be surveyed in the SMC will be $3\times$ larger than that covered by Massey \& Duffy (2001).

The choice of telescopes and instruments was straight-forward.  As far as the imaging was concerned, one thing we had learned from the Massey \& Duffy (2001) survey was that good image scale is crucial in dealing with the crowding that often characterizes the location of massive stars in the Magellanic Clouds.  Thus, we would rather deal with more fields (and a longer project) than accept a pixel scale of 1-2\arcsec.   At the same time, the larger the field of view (FOV), the better in keeping the project finite.  We realized that the Las Campanas 1~m Swope would provide a modestly large FOV (0.094 deg$^2$) with good image scale (0\farcs435 pixel$^{-1}$).  Despite this, it would require $\sim 450$ fields to cover the LMC, and $\sim 330$ for the SMC, allowing for some overlap between fields.  We estimated that we would achieve
adequate signal-to-noise in a few minutes per filter, and thus if we planned on 6 fields per hour (optimistic as it turned out) the survey would require ``only" 130 hours, or about 20 nights of observing, a large but not impossible amount. (The camera has since been replaced with an even better system,
with a 0.21 deg$^2$ FOV, a factor of 2.2$\times$ larger, more than off-setting our optimism about the number of fields per hour; see below.)   As for the spectroscopy, the faintness ($V\sim 16$) and need for good signal-to-noise at reasonable exposure times led us to the Magellan telescopes, and in particular the Magellan Echellette Spectrograph (MagE) implemented on the 6.5~m  Clay telescope (Marshall et al.\ 2008).

 \subsection{Imaging}
 
 Our recent WR surveys of M33 (Neugent \& Massey 2011) and M31 (Neugent et al.\ 2012a) used the ``standard" {\it WC}, {\it WN}, and {\it CT} WR filters developed and first used by Armandroff \& Massey (1985). These filters have bandpasses that are $\sim$50~\AA\ wide, and are centered on C III $\lambda 4650$ ({\it WC} filter), He~II $\lambda 4686$ ({\it WN} filter), and continuum at $\lambda 4750$ ({\it CT} filter).   
 We needed new versions of these that would fit the 3-inch $\times$ 3-inch filter holder of the Swope, and we had 2-cavity interference filters made to similar specifications by the Andover Corporation.   The filters
have very good transmission ($\sim$80\% at their peaks) and full-widths at half-maximum (FWHM) bandpasses of 51-55~\AA, with 10\% band widths of $\sim$90~\AA.  We note that one problem with this filter set is that foreground M dwarfs can be picked up as WR candidates as they have a strong TiO absorption band at $\lambda 4760$, which falls in the continuum band.  We were confident, however, that 2MASS colors would allow us to eliminate the majority of such spurious candidates in the Magellanic Clouds, an option we did not have in the case of M31 and M33, as those candidates were much fainter than sources in the 2MASS catalog.
In Fig.~\ref{fig:test} we superimpose the transmission curves of these filters on two fluxed spectra of newly discovered WR stars we will discuss later in this paper.  
  
 Our inaugural imaging run took place on 9 nights 2013 Sept 21-29 (UT) using the SITe\#3 CCD camera.   The CCD was a $2048\times4096$ device (formatted to $2048\times3150$ to avoid vignetting) with 15$\mu$m (0\farcs435) pixels; each exposure thus covered 14\farcm8$\times$22\farcm8.  On the first night we observed 18 fields with
120 s exposures through each of our filters.   The seeing was poor, 1\farcs6-3\farcs0.  Examination of the data in real time cast doubt that we were going sufficiently deep: we found that we were detecting the faintest and weakest-lined of the known WNs but not with extremely high confidence.  After the first night we increased our exposure times by a factor of 2.5 (i.e., by 1 mag) to 300 s per filter, and repeated all of the first night fields with these longer exposure times, finding that we were now extremely sure of our detections for all of the previously known WRs.  We were closed due to high winds and poor seeing for most of the second night, and had some unusually poor seeing and clouds on the sixth night, but for the remainder of the time we had good conditions.  Our typical seeing was 1\farcs2-1\farcs9, with a median value of  1\farcs5.  Occasionally the seeing was worse than 2", in which case we repeated the field on another occasion.  

The observing run substantiated that our goal of covering all of the Magellanic Clouds is practical, and we can now provide a more accurate estimate of what will be needed.  In addition to 15 minutes of exposure per field, our overhead was significant, and dominated by the relatively long read-out times, 127 s per image. Unfortunately we discovered that we could not slew the telescope to the next target during the final readout without introducing streaks into our data, and so the total time spent on reading out the chip was 6.4 minutes per field.  Slewing and setting up the guider did not take much time, usually just 2-3 minutes per field, and so in practice we managed about 2.5 fields per hour. 
We also stopped to focus and check the telescope
pointing 2-3$\times$ a night, each requiring about 10 minutes.   In the end we obtained satisfactory data on 51 fields in the SMC, and 76 fields in the LMC (127 in total), about 15-17\% of our desired coverage for each galaxy, in essentially 6 good nights (eliminating the times of poor seeing, high winds and clouds) observing about 8.5 hours on the Clouds each night.  With the new camera, we will have a $2.2\times$ improvement in areal coverage and much shorter readout times, and we foresee another $\sim$15 clear nights will be needed to complete the survey. We have been fortunate in already being assigned 10 nights for the 2014 Magellanic Cloud observing season. 

We took $\sim$10 bias frames and 3 dome flats per filter daily, 
and tried to obtain good sky flats through at least one of the filters
each evening.    We reduced the data nightly using IRAF\footnote{IRAF is distributed by the National Optical Astronomy Observatory, which is operated by the Association of Universities for Research in Astronomy (AURA) under cooperative agreement with the National Science Foundation.}. We found that the biases were indeed useful, as 
there was a 12 ADU turn-up on the left side of the images that was well taken out by the bias frames.   
The dome flats did an excellent job of removing the pixel-to-pixel variations and large-scale donuts due to dust specks on the dewar window, but the sky flats were necessary to remove a 4-5\% gradient in the y (NS) direction due to the dome flats not illuminating the chip quite the same way as the sky.   

Prior to running the image subtraction software, we needed to carefully align the images.  In order to determine accurate shifts we used IRAF's {\it daofind} to find sources that were 10$\sigma$ or higher above the noise of the background.   We then matched the sources between the three filters with the assumption that the shifts were small, finding the median
pixel shifts.   Next, we shifted the {\it WN} and {\it WC}  images to align with the {\it CT} images using a cubic spline interpolation.  We also added an accurate world coordinate system to the headers of each shifted image using the ``astrometry.net" software (Lang et al.\ 2010).  At this point the images were ready for image-subtraction.

For the image subtraction, we used the High Order Transform of PSF And Template Subtraction ({\it HOTPANTS}) code written by Andrew Becker, described briefly in Becker et al.\  (2004) and in more detail on his webpage\footnote{http://www.astro.washington.edu/users/becker/v2.0/hotpants.html}, as Carlos Contreras (2013, private communication) reported excellent results with this code on images taken with the Swope.  Fig.~\ref{fig:wow} shows the results of the image subtraction on one of our SMC fields.  There are two known WRs in the image, SMC-WR6 (AzV 332 = Sk 108) on the left, and SMC-WR12 on the right.  The former is the brightest ($V=12.3$) WR in the SMC (other than HD 5980, an unusual LBV/WR object), while the latter is the faintest ($V=15.5$) and among the weakest-lined WRs;  see Table 1 in Massey et al.\ (2003).   Both WRs show up unambiguously in the difference frame at the bottom.  The bright WR is not close to saturation (on our {\it WN} frame it is 1.5 mag fainter than where non-linearity would become a significant issue), while the faint WR has a nominal signal-to-noise on the resultant image of 200.  

After the subtracted images were produced, each of the resultant images was examined for WR candidates.  All previously known WRs in our frames were readily detected except for the most crowded members of R136.   Candidates were classified into three classes depending upon how certain we felt about the detection and the significance of the magnitude differences we computed from photometry of the frames.   The list was then checked against known
objects in the Clouds and 2MASS photometry; this allowed us to eliminate many planetary nebulae (which will
show He~II $\lambda 4686$ nebula emission if the radiation field is hard enough), known Of-type stars, and very red stars which we expect to show up because of the TiO absorption feature in the {\it CT} filter.   

In practice the process of identifying candidates from the subtracted images was somewhat iterative.  We had an opportunity to observe a few candidates spectroscopically within a few weeks of our imaging run, and selected 8 candidates that spanned a range in our confidence level to help us evaluate our search. We were greatly encouraged that most but not all of these proved to be new WR stars (as described below), 
substantiating that our project was not in vain.

\subsection{Spectroscopy} 

For our spectroscopic followup, we used MagE (Marshall et al.\ 2008) on the Clay 6.5-m Magellan telescope.
We had previously been assigned two nights, UT 2013 Oct 16 and Dec 14, 
as part of our spectroscopic survey of massive stars
in the Magellanic Clouds (see, e.g., Neugent et al.\ 2012b).  
Francesco Di Mille kindly provided a few additional spectra
obtained on UT Oct 18 and Dec 16, 2013, during engineering time.
Exposure times were typically 600 s, and were followed by a 3 s Th-Ar exposure to provide wavelength calibration. 

MagE provides wavelength coverage from the atmospheric cut-off ($\sim$3200~\AA) through 1$\mu$m at a resolving power of 4100 with a 1" slit.   
Because of the large wavelength coverage, obtaining flat fields with sufficient
signal-to-noise so as to not degrade the actual spectra is very tricky.  We have found that we can achieve high signal-to-noise (100-200) by {\it not} flat fielding, owing to the intrinsic flatness (uniformity) of the chip (see discussion in Massey \& Hanson 2013).   The spectra were extracted using Jack Baldwin's ``mtools" IRAF routines, and wavelength calibrated
and fluxed with the usual IRAF echelle reduction tasks.
Spectrophotometric standards were observed in order to remove the blaze function of each order and to provide flux calibration.  Further details of our reduction procedure can be found in Massey et al.\ (2012).

During the October observing we confirmed 5 new WRs in the LMC; the other 3 candidates we observed were planetary nebulae (PNe) or red stars.  
By the December observing date we had completed our inspection of the remaining fields, and were prepared with a list of 7 priority 1 candidates,  25 priority 2 candidates, and 13 priority 3 candidates.  All of the priority 1 and 3 candidates were in the LMC, while 9 of the 25 priority 2s were in the SMC.  Other candidates had been eliminated by literature searches as either being Of-type stars, PNe, or red stars.   Of the 7 priority 1s, 3 proved to be Of-type stars, and the other 4 newly found WRs, bringing the total number of newly found WRs to 9.  None of the priority 2s or 3s proved to be WRs, although some proved to be Of-type stars.  We describe these and the other interesting stars in the following section.  During the December observing time we also re-observed the five new WRs we had found in October, as all had shown absorption lines in addition to emission, and we wanted to further investigate the nature of these stars.

\section{Spectral Classifications}
\label{Sec-results}

Our study has identified 9 new WNs in the LMC, along with 5 previously unknown Of-type stars in the LMC plus one in the SMC.  Two of these turn out to be members of the  ``Of?p" class of magnetic O stars. In addition, we found one early-type O star (not of Of-type) accidentally.  We re-observed one known B[e] star, and were struck by the presence of broad He~II $\lambda 4686$ emission; we argue below this may be a B[e] + WN binary.  We had no difficulty recovering most of the previously known WRs in our LMC fields (119 WRs) and SMC fields (12 WRs);  the only exceptions were the very crowded WRs in the R136 region.   We also ``rediscovered"  9 known Of stars along with several PNe.

We list the newly confirmed WRs
in Table~\ref{tab:WRs} and the other new discoveries in Table~\ref{tab:Ofs}.   We provide cross references to
Massey (2002), who gives {\it UBVR} photometry for many of the massive stars in the Magellanic Clouds,
along with cross references to the near-IR catalog of Kato et al.\ (2007).  The Kato et al.\ (2007)  survey goes a few magnitudes deeper than 2MASS, but more to the point has considerably greater spatial resolution. We refer to these stars by our field designation i.e., LMC172-1 happens to be the first star identified in field 172 in the LMC.  When our survey is complete we plan to provide a more rational designation, but for now these serve as useful short identifiers.  

The coordinates listed in the table are on the ICRS system and are good to a fraction of an arcsecond.  We are indebted to Brian Skiff for helping us refine these.  We have provided finding charts in Fig.~\ref{fig:FCs} for the five stars that were crowded; the rest should be easily
identified from their coordinates. 

We include our {\it CT} magnitude, the zero-point of which has been set from the $V$-band magnitudes of the known WRs.  We do not expect this to be particularly accurate, but in nearly all cases we also have actual $V$-band magnitudes from either Massey (2002) or Zaritsky et al.\ (2004), and these usually agree to within 0.1~mag with the {\it CT} mag.  We have computed the absolute magnitudes assuming a distance of
50~kpc for the LMC and 59~kpc for the SMC (van den Bergh 2000), corresponding to true distance moduli of 18.5~mag and 18.9~mag, respectively.  Except for LMC 174-1, the faintest star in our sample (with $V=17.2$) , we adopt an average extinction $A_V$ of 0.4~mag and 0.3~mag, respectively, for the LMC and SMC stars, based upon Massey et al.\ (1995).  These values are consistent with the reddenings we infer from our fluxed spectra, although we note that modeling would be useful for improved estimates of the reddenings.  For LMC 174-1, the fluxed spectrum indicates an additional 1.2~mag of extinction, as noted in Table~\ref{tab:WRs}.

\subsection{Spectral Classifications: Newly Found WRs}

We classified the WRs following the same premises as in our M33 and M31 studies, i.e., using the criteria originally introduced by Smith (1968a, 1968b) and extended to earlier and later types by van der Hucht et al.\ (1981).  
Before discussing the stars individually we need to comment on
one subtlety of the classification of early WNs.   Many of the previously known WNs in the LMC are of WN3 subtype (BAT99), in which N~IV $\lambda 4058$ is ``much weaker" than N~V $\lambda \lambda 4603, 19$.    The only earlier type defined at the time of BAT99 was WN2, in which N~V (along with N~IV) is absent.  Van der Hucht (2001) introduced the intermediate WN2.5 type in which N~V is present but N~IV is absent, at some unspecified signal-to-noise.
If we were to classify our 8 WNs in that way
they would all be of WN2.5 type.  However,  examination of our old spectra (Conti \& Massey 1989) of 
LMC WRs classified as WN3 confirms that these too would be classified as WN2.5.  
So, for consistency with older works, we eschew the WN2.5 class and refer to these stars as WN3.

\subsubsection{WO! Another One!}

\paragraph{LMC195-1: WO2.}   This star is a rare find, a WO-type Wolf-Rayet, only the third known in the LMC.  Its spectrum is shown in the upper panel of Fig.~\ref{fig:LMC1951}.  WO subtypes are based primarily on the ratio of
the O~VI $\lambda \lambda 3811,34$ to O~V$\lambda 5590$ (Crowther et al.\ 1998). We measure an equivalent width ratio of 4.5, making this an WO2.   

The star is located within the LH41 association, home to S Dor and R85 (two LBVs), Br 21 (B1 Ia+WN3 star), and numerous O stars and B supergiants and even a rare F-type supergiant.  (See Table 1 of Neugent et al.\ 2012b.) 
As can be seen from the finding chart in Fig.~\ref{fig:FCs}, our new WO2 star is just 9\arcsec\ north of LH41-1042, the WO4 star we discovered two years ago (Neugent et al.\ 2012b). We were concerned for a moment that we had possibly observed the wrong star and reobserved LH41-1042,  but we can reject this for three reasons.  First, we made careful use of a finding chart at the telescope.  Second, the telescope coordinates of the spectra of the objects observed both before and afterwards show small consistent offsets with the intended coordinates, and the observation of LMC195-1 shows the same offset to better than 1".  And third, although both stars are classified as WO, their spectra are not identical by any means (hence the difference in the WO subtype).  For comparison, we show the spectrum of LH41-1042 in the lower panel of Fig.~\ref{fig:LMC1951}.

\subsubsection{New WNs}

Our eight other newly found WRs are all WN3 stars.  All show absorption lines in addition, but in only two cases are we convinced that these are likely binaries.

\paragraph{LMC079-1, LMC170-2, LMC172-1, LMC199-1, and LMC277-2: ``WN3+O3~V."} The spectra of these 5 stars are essentially indistinguishable, and are shown together in Fig.~\ref{fig:O3s}.  We have selected a star (LMC 277-2) with high signal-to-noise data to illustrate the blue part of the spectrum in Fig.~\ref{fig:O3}.   The stars have strong N~V $\lambda\lambda 4603,19$ and He~II $\lambda 4686$ emission,  and exhibit absorption lines characteristic of early O-type. We classify the WR components as WN3, as no N~IV $\lambda 4058$ is present, although N~V $\lambda 4945$ emission is present.   He~II $\lambda 6560$/H$\alpha$ is also strongly in emission.   The absorption spectra consist of Balmer lines and He~II.  Despite our good signal-to-noise (60-120 per 1~\AA\ resolution element), there is no trace of He~I $\lambda 4771$ or weaker He~I lines, and so we classify the absorption as O3~V\footnote{Walborn et al.\ (2002) extended the O-type classification to the O2 type based upon the relative strengths of N~IV $\lambda 4058$ and N~III $\lambda 4634,42$.  Since none of the stars here have either line, we do not attempt to distinguish O2s from O3s; we instead use ``O3" inclusively.}.  The lack of emission from
Si IV $\lambda \lambda 4089, 4116$ or N~IV $\lambda 4058$ argues for the dwarf luminosity class, while the spectral subtype is due to the lack of He~I and strong He~II $\lambda 4200$ and $\lambda 4542$ absorption. 

All five stars have absolute visual magnitudes that are fainter than either WN3s or O3~Vs as we argue further in Section~\ref{Sec-sum}, where we more fully discuss the nature of these objects.  Here we simply note that  we obtained a second observation for LMC079-1, LMC170-2, and LMC172-1 in order to see if the radial velocities of the emission and absorption lines varied in anti-phase. Of course, without three or more observations it is hard to evaluate our measuring errors, particularly with very broad emission lines and relatively weak absorption.  We measured the radial velocities both by using the line centroids and by using cross-correlation.  In the end, our results
were  ambiguous, as the agreement in the velocity shifts was not very consistent between the two methods.  Although we found small velocity shifts for these three stars, there was no indication that the absorption and emission moved in opposite senses.  Additional radial velocity monitoring is planned for the next observing season.  
 
 \paragraph{LMC143-1: WN3+O8-9 III.}  The spectrum is shown in Fig.~\ref{fig:LMC1431}.  We find an emission-line spectrum characteristic of an early WN-type WR plus the absorption spectrum of a mid-to-late O-type star.
Strong N~V $\lambda\lambda 4603,19$ and He~II $\lambda 4686$ are in emission, along with He~II $\lambda 4542$,  N~V $\lambda 4945$, He~II $\lambda 5412$, and He~II $\lambda 6560$/H$\alpha$ emission. The latter is split into a double peak by an absorption component. There is no N~IV $\lambda 4058$ nor N~III $\lambda \lambda 4634,42$ emission, and so again we classify the WR component as WN3.  The absorption spectrum is dominated by the Balmer lines and He~I, with strong He~I $\lambda 4387$ and $\lambda 4471$.  He~II $\lambda 4200$ may be barely present in absorption.   Si~IV~$\lambda 4089$ is modestly in absorption but there is no sign of Si III $\lambda 4553$ despite our S/N of 150 per 1~\AA\ spectral resolution element.  The presence of He~II emission makes exact classification of the O star uncertain, but given the strength of Si IV and lack of Si III we conclude the O star is roughly O8-O9 III.  The giant classification follows from Si IV $\lambda 4089$ being half as strong as He~I $\lambda 4026$.   

We also obtained two observations of this star.  The emission and the absorption line velocity shifts are anti-correlated, as one expects for a double-lined binary, although the change is somewhat marginal compared to our estimated errors.  
 
\paragraph{LMC173-1: WN3+O7.5~V.}  We illustrate the spectrum of this star in Fig.~\ref{fig:LMC1731}.  We again see an emission spectrum typical of an  early-type WN  plus the absorption component of a mid-O-type star. There is emission at N~V $\lambda \lambda 4603, 19$ and He~II $\lambda 4686$, as well as  N~V $\lambda 4945$, He~II $\lambda 4860$/H$\beta$, He~II $\lambda 5412$,  and He~II $\lambda 6560$/H$\alpha$.  The absorption spectrum is that of an intermediate O-type, with
 He~I $\lambda 4471$ being just a bit stronger than He~II $\lambda 4542$ and modest He~I $\lambda 4387$ being
 present.  We classify the O star as O7.5.  N~III $\lambda 4511-17$ absorption is present, as we would expect for
 an O7.5~V (e.g., see the spectrum of HDE~319703A illustrated in Sota et al.\ 2011), as well as modest N~IV $\lambda 4058$ absorption.  The apparent lack of any N~III $\lambda \lambda 4634, 42$ emission argues that the star is a dwarf, which is consistent with its $M_V$.
 
 We obtained two observations of this star, and in this case the absorption and emission radial velocities were clearly anti-correlated,
 with a shift of $\sim$ -100 km s$^{-1}$ for the absorption and +250 km s$^{-1}$ for the emission.  

\paragraph{LMC174-1: WN3+ early O.}  The spectrum of this star is shown in Fig.~\ref{fig:LMC1741}.  Its spectrum is very similar to that of the ``WN3+O3~V" stars discussed above, but is described separately here as the WR emission-line spectrum dominates, with only weak absorption present at H$\delta$ and He~II $\lambda 4200$ and $\lambda 4542$.  
The WR component is again a WN3, with N~V $\lambda \lambda 4603, 19$, He~II $\lambda 4686$, N~V $\lambda 4945$, He~II $\lambda 5412$, and He~II $\lambda 6560$/H$\alpha$ all present.  Nebular emission that was not completely subtracted is apparent at H$\beta$ as well as 
[OIII] $\lambda \lambda 4959, 5007$, H$\alpha$, [NII] $\lambda 6584$ and [SII]  $\lambda \lambda 6717, 31$.
If the O star is late enough to have He~I $\lambda 4471$ it is filled in with emission.  We classify the system as WN3 + early O.  Of the stars in the sample, it is the only one whose fluxed spectrum indicates significant reddening, with 1.2~mag more extinction than that of the other stars.  Applying this correction leads to $M_V\sim -3.0$, very similar to that of the ``WN3+O3~V" objects discussed above.

\subsection{Other Interesting Stars}

As we have emphasized in our previous surveys for WRs, a critical test of completeness is whether or not the sensitivity is sufficient to detect even Of-type stars.  We discuss this further in Section~\ref{Sec-completeness}, but here we are heartened by the fact that we not only recovered known Of-type stars but discovered new ones.  Two of these are newly found members of the ``Of?p" class
(see, e.g., Walborn et al.\ 2010), while four others are normal Of-type supergiants.  Our spectroscopy also accidentally found an early-type O4~V star.  Finally, we comment upon a previously known B[e] star suggesting it might have a WN-type companion.   All the stars discussed in
this section are listed in Table~\ref{tab:Ofs}

\subsubsection{Two O8f?p Stars}

 \paragraph{SMC159-2 and LMC164-2: O8f?p.}   
 The spectra of these two stars are shown in Fig.~\ref{fig:Ofs} in comparison with the Of-type supergiants discussed below.
 Both of these stars have very strong He~II $\lambda 4686$ emission, along with
 much weaker N~III $\lambda \lambda 4634,42$ and C~III $\lambda 4650$ emission.   These emission line signatures
 are characteristic of Of?p objects, a class introduced by Walborn (1972) to describe the peculiar spectra of
 the Galactic O stars HD~108 and HD~148937, which show very strong He~II $\lambda 4686$ emission relative to that
 of N~III $\lambda \lambda~4634, 42$, and C~III $\lambda~4650$ emission that is comparable to N~III. 
 Subsequent studies have shown that this class likely consists of magnetically-braked oblique rotators 
 (see discussion in Walborn et al.\ 2010).   Both SMC159-2 and LMC164-2 are likely members of this class\footnote{We are indebted to our referee, Nolan Walborn, for suggesting that the Of?p classification should apply to these two stars.}.  
Both stars show He~I $\lambda 4471$ absorption just a bit stronger than that of He~II $\lambda 4542$, making these both O8f?p.
SMC159-2  has a He I profile that is considerably broader than that of He II, while the reverse appears to be true for LMC164-2; this is
understandable given that He I may have a circumstellar component in Of?p stars.   Finally, both
stars show narrow emission superposed on the lower Balmer absorption lines.  These do not appear to be nebular as [OIII] is not present.
There is no nebulosity visible around SMC159-2 in the digitized sky survey, and only a little around LMC164-2.   
Inspection of the two-dimensional spectra confirms that that the emission is not spatially extended around SMC159-2. In the case of LMC164-2, it is a little harder to tell as there are faint nebular lines present, including [O~III], but the fact that [O~III] subtracted out well in the reductions again suggests a local original for the Balmer emission.  We conclude that the Balmer emission is circumstellar.  An alternative classification
of these stars as supergiants is ruled out by the lack of Si~IV $\lambda 4089$ absorption. 

\subsubsection{Of-type Supergiants}

  \paragraph{LMC104-2: O3.5~If + O6-7.}   Strong N~IV $\lambda 4058$ and N~III $\lambda \lambda 4634,41$  emission are apparent, with
 He~II $\lambda 4686$ showing a P Cygni profile.  Strong He~II absorption is present along with He~I $\lambda 4471$ with He~I $\lambda 4387$ only very weakly present.   We would
 classify this as an O3.5 If + O6-7 pair.  The O3.5 If classification comes about due to the similarity in strengths of the N~IV and N~III emission (Walborn et al.\ 2002).  The strong Si IV $\lambda \lambda 4089, 4116$ emission combined with N~IV $\lambda 4058$ emission argues that the star is an early O supergiant.   The
 presence of He~I $\lambda 4471$ combined with the weak presence of $\lambda 4387$ would suggest an O6-O7
 star is also present.    The blue absorption component of He II $\lambda 4686$ is likely part of a P Cyg profile (typical of O3.5 If stars), rather
 than signifying that the companion is a dwarf.
    
 \paragraph{LMC156-1: O6 If.}  This star has modest N~III $\lambda \lambda 4634,42$ and He~II $\lambda 4686$ emission.  In combination with the absorption spectrum, we classify it as O6 If. 
 
\paragraph{LMC173-2: O7.5 Iaf.} The star is clearly an Of-type star, and not a WR.  Weak N~III $\lambda \lambda  4634, 41$ and He~II $\lambda 4686$ emission is present, but no other emission lines are seen. The spectrum is readily classified as O7.5 Iaf.

 \paragraph{LMC174-4: O4 Ifc.}  This is another Of-type star, with N~IV $\lambda 4058$, Si IV $\lambda \lambda 4089, 4116$, N~III $\lambda \lambda 4634,42$, C III $\lambda 4650$, He~II $\lambda 4686$, and H$\alpha$ emission.   He~I $\lambda 4471$ is readily discernible despite it having an equivalent width of 85~m\AA, thanks to our high S/N (200) spectrum.  We classify the star as O4 Ifc, with the ``c" 
 due to the strong C III component.  For comparison, see, e.g., the spectrum of CPD -47 2963 illustrated in Figure 3 of Sota et al.\ (2011).

\subsubsection{An Accidental Find of an O4~V Star}

 \paragraph{LMC174-3E: O4 V.}  We ``rediscovered" the Of-type star [ST92] 5-31 as part of our survey, calling it LMC174-3.  This star was first classified by  Testor \& Niemela (1998) as one of the very rare O3 If* stars, and more recently reclassified with the advanced notation   ``O2-3(n)f*p" by Walborn et al.\ (2010).  Owing to some confusion with the cross-identification and with the finding chart, we wound up not only re-observing this star, but also the fainter star located 3" to the east, which we designate LMC174-3E.  The fainter star is not marked on the finding chart of Testor \& Niemela (1998), but the isophotes do show [ST92] 5-31 as elongated east and west.  The NIR survey of Kato et al.\ (2008) identifies multiple sources at this position.
 We classify it as O4~V.  
 
 \subsubsection{A B[e]+WN Binary?}
 
 \paragraph{LMC174-5 = HD 38489 = LHA 120-S134 = Hen S 134: B[e]+WN?}  This star is often referred to as a B[e] star (e.g., Zickgraf et al.\ 1986), and was likened to $\eta$ Car, S Dor and other LBVs by Shore \& Sanduleak (1982) and van Genderen (2001).  The star is discussed extensively by Shore \& Sanduleak (1983), and a high dispersion photographic spectrum is shown and briefly discussed by Stahl et al.\ (1985)---see their Fig.~29.  Technically B[e] stars should show absorption features
 typical of a B star but in fact absorption has not been seen in this star, although as Conti (1997) remarks such absorption is often weak and difficult to detect.  We ``rediscovered" this star, and thought it would be useful to take a modern spectrum of it, as the previous work has mostly been photographic.   Our spectrum is similar to the one shown by Stahl et al.\ (1985),  with strong Balmer emission and [FeII] and Fe~II lines, similar to what we observe in AE And and other ``hot" LBV candidates in M31 and M33; see Figs.~10-12 of Massey et al.\ (2007).  However, what we found most intriguing was the  broad He~II $\lambda 4686$ feature.  The relevant section of our spectrum is shown  in Fig.~\ref{fig:HD38489}.  Although B[e] stars often display broad stellar wind features in addition to sharp emission lines (Zickgraf et al.\ 1985, 1986),  a broad He~II $\lambda 4686$ feature is unique amongst such objects as far as we know.  This feature was mentioned as a curiosity by others, e.g.,  Shore \& Sanduleak (1983), Stahl et al.\ (1985), and Zickgraf et al.\ (1986).  We propose an alternative explanation, namely that this star is a B[e]+WN binary. We note that the star is an x-ray source (appearing in both the ROSAT All-Sky Bright Source Catalogue and XMM-Newton Serendipitous Source Catalog), unlike other bright Magellanic Cloud B[e] stars (i.e., R126 = HD 37974 = Hen S127). At $V\sim$12.0, this star serves as an example of how superficially we know the massive star content of our nearby extragalactic neighbors.  Clearly a modern study of this star is warranted. 

\section{Completeness}
\label{Sec-completeness}

The critical issue surrounding all searches for WRs is that of completeness: if the goal is to compare the number ratios of WCs and WNs, then being complete for the weaker-lined WNs is necessary.  Such surveys are mostly {\it flux-limited}, but not entirely: a bright star with a small equivalent width will still have a substantial line flux but might be hard to detect because of the low contrast between the on-line and off-line exposures.  A very faint star with a large equivalent width will have a small line flux but might be easily detectable from the high contrast, as long as the survey is sufficiently deep.  In our case,
6 of our 9 newly discovered WRs are faint ($M_V\sim-3$) and have weak emission-line strengths, 
sadly calling into question all previous ``complete" surveys.  Of course, to keep this in perspective, we are talking about the addition of 6 unusually faint WRs compared to 131 previously known in the same fields, a 4\% issue, and within the 5\% completeness limit that we have estimated for our M33 and M31 surveys (Neugent \& Massey 2011, Neugent et al.\ 2012a). 
Still, it will be interesting to see how this plays out as additional area is surveyed and how much the WC/WN ratio changes in these galaxies.

We can address the completeness issue somewhat more quantitatively by using photometry from our images to compare the ``detectability" of the newly found WRs with previously known WRs and the Of-type stars.  In the upper two panels of Fig.~\ref{fig:completeness}
we plot the {\it CT} magnitude vs.\ the {\it WN}$-${\it CT}  or {\it WC}$-${\it CT} magnitude differences for WN and WC stars, respectively.  We have separated the plots for the LMC and SMC as the  SMC WNs are often described as the ``weakest-lined" WRs known (see, e.g. Massey \& Johnson 1998).    We are intrigued to find that the distinction still holds.  Although the newly found WNs in the LMC have weak emission (in the sense of their not having
largely negative {\it WN}$-${\it CT} values and are fainter on average), the SMC WNs are in fact closer
to the {\it WN}$-${\it CT}=0 line.  There are also other WRs already known in the LMC which are equally weak-lined and faint.  

Given this, why have our new WRs not been discovered before now?  In Section~\ref{Sec-intro} we emphasized the fact that many of the known WRs in the LMC were found accidentally--the result of spectroscopy rather than as part of systematic searches.  We believe this underscores the necessity of the present survey.

We can examine these data another way.  We have made the point above that the detectability will depend not only on
the emission line fluxes but also on the continuum magnitude.  One way of combining these two is by considering the ``significance level" of the magnitude differences.  This was first used by Armandroff \& Massey (1985), and discussed further by Massey \& Johnson (1998).  If we detect a certain magnitude difference between the on-line WR filter and the off-line continuum filter, how significant is this difference compared to the photometric error associated with the difference?  In other words, if the magnitude difference is $-0.5$~mag and the uncertainty in the magnitude is 0.05~mag, we would consider this a 10$\sigma$ detection.  This is probably what limits the detectability when we visually examine the difference frames after our image subtraction.  If the star is faint and the magnitude difference is small, we are less likely to detect the object, particularly if the residual image is comparable to the noise on the subtracted frame. Such a situation would result in a low significance level.  At the same time if a brighter star has the same magnitude difference, the residual image will rise above the noise.  Of course, the significance level will depend upon the details of the survey and the observing conditions: a 3$\sigma$ detection with one set of exposure times could be a 30$\sigma$ detection if the exposure time were increased a hundred fold, although the magnitude difference would remain the same.  One should recall that we adjusted our exposure times to make the faintest and weakest-lined WNs known in the SMC easy detections.

We show the ``significance" plots in the lower half of Fig.~\ref{fig:completeness}.  We have had to use a log scale, as there are known WRs which have significance levels of over 100$\sigma$!   We note that all of the newly found WRs have significance levels greater than 10$\sigma$.  Yet, Of-type stars with much lower significance levels were readily detected.  We believe this strongly argues that our survey is finding what we set out to find.

In both kinds of plots, the Of-type stars represent an extreme: their emission-line equivalent widths are closer to zero than those of most WRs, but they are also considerably brighter on average.  We see that they occur at lower significance levels than do the vast majority of our WRs. 

We do note that we failed to find several of the known WRs under the extremely crowded conditions of the R136 cluster.  This region is unique within the Magellanic Clouds (and indeed within the nearby universe) and we are satisfied that under ``typical" crowding conditions we are complete, as shown, for instance,  by the discovery of the new WO2 star in a crowded knot in LH-41.

\section{Discussion: the Nature of our Discoveries}
\label{Sec-sum}

We have described the first exciting results of our survey.  Despite having covered only 15\% of the LMC and SMC,  we have confirmed 9 new WR stars in the LMC (an increase of 6\%), and suggested that a well-known B[e] star, HD 38489, may be a tenth.  
We have also identified 2 of the rare Of?p objects, 4 previously unknown Of supergiants, and an O4 dwarf.  
We detected all of the known WRs in these fields (except the most crowded members of R136 in 30 Dor), as well as many previously known Of-type stars. We have argued that our survey is going both faint enough and that our detection method is sensitive to even the weakest-lined WRs.

The most remarkable aspect of our find is not the quantity of new WRs, but their characteristics.  First, one of the newly found WRs is a WO star, only the third to be found in the LMC.  It is of WO2 type, and is located just 9\arcsec\ (2.2 pc in projected distance) from the WO4 we found two years ago (Neugent et al.\  2012b).  The other 8 newly found WRs are WN3s that also show absorption lines. Two of these WN3s appear to have normal mid-to-late O-type companions and show radial velocity variations consistent with a binary nature.  However, our most remarkable find has been that of the five stars we would naively classify as ``WN3+O3~V."

The presence of absorption in the spectrum of a WR star is nearly always indicative of binarity.  If absorption is otherwise present, it is usually combined with P Cygni emission, such as the case for the very luminous and massive hydrogen-rich late-type WNs seen in the R136
cluster (Massey \& Hunter 1998, Crowther et al.\ 2010) and in NGC 3603 (Melena et al.\ 2008).  Those are unevolved stars but which are
so luminous that their winds mimic the emission found in evolved WRs.    Possibly a closer analogy to our ``WN3+O3~V' objects are some of the SMC WN3 stars which show the absorption signature of an
early-type O star, although none as early as O3~V (Table 1 of Massey et al.\ 2003). None of these have been shown to be binaries.
However, they are all significantly more luminous than ours ($M_V\sim -3.6$ to $-5.5$) and therefore could simply be multiples viewed at unfavorable inclinations, or whose components are too widely separated to have detectable radial velocity variations.

What, then, is the nature of our ``WN3+O3~V" objects? 
There are several reasons why these stars are unlikely to be actual WN3+O3~V pairs.  First, O3~V stars are the ``rarest of the rare," as only the most luminous and massive stars start their lives in an O3~V phase. (Only stars of 50$M_\odot$ and larger obtain sufficiently high effective temperatures to be spectroscopically identified as O3 stars; see, e.g., Ekstr\"{o}m et al.\ 2012.) Thus, outside of the concentration of O3 stars in the very young and massive R136 cluster (Massey \& Hunter 1998), only about a dozen O3~V stars are known in the entire LMC (Skiff 2014).  So, to have come across five O3~V stars that just all happen to be members of a binary system with WN3 stars seems rather far-fetched.  A second, and perhaps more irrefutable argument, is that the absolute magnitudes of these ``WN3 + O3~V" systems are all quite faint, with $M_V=-2.3$ to $-3.0$.  But,
this is much fainter than an O3~V star ($M_V\sim -5.4$, Conti 1988), and in fact is even faint for a WN3 
($M_V\sim -3.8$, Hainich et al.\ 2014).    Thus, there would seem to be no way that these objects can truly consist of a WN3+O3~V pair. Third, we have two observations for three of these systems, and none show the radial velocity variations we might expect to find  in binaries.
Finally, such a WN3+O3~V system would be very hard to understand from an evolution point of view: the O3~V component must be quite young ($<$1-2~Myr), while it would have taken several million years to have formed the WR component.

With five such objects (and likely a sixth) we are forced to conclude that we have discovered a hitherto unrecognized class of WRs,
stars that are under-luminous visually and whose winds are thin enough to show underlying absorption. For absorption lines to be present from the WR itself requires a different set of physical conditions in the stellar wind than is found in other WRs.  Are these ``WN3+O3~V" stars even evolved objects?  
A preliminary effort at modeling the optical data of LMC170-2 using CMFGEN (Hillier \& Miller 1998) shows that a  good 
match to the observed spectrum (emission {\it and} absorption) can be achieved with a model using a high effective temperature ($\sim$80,000-100,000 K) along with a strongly enhanced helium (He/H$\sim$1.0 by number) and nitrogen abundances ($\sim$10 $\times$ solar),
indicative of advanced CNO processing.   The models require mass-loss rates of  $0.8-1.2\times 10^{-6} M_\odot$ yr$^{-1}$, corrected for clumping using a volume filling factor of 0.1.    Fig.~\ref{fig:john} shows how successful the best-fitting model matches the spectrum. The high effective temperatures would imply a bolometric luminosity of $\log L/L_\odot\sim$ 5.3-5.6.   These physical parameters are all in accord with what we expect for LMC WN3 stars  (Hainich et al.\ 2014), except for the mass-loss rate, which is lower than what we expect for a WN3 star by a factor of 3 (see Fig.~6 of Hainich et al.\ 2014),  and more similar to what we would expect from O2-3~V stars (see, e.g.,  Massey et al.\ 2005).  
However, none of these values are well determined by the optical data alone, as we lack lines arising from multiple ionization stages of the same species, severely hindering our ability to constrain the effective temperature.  For instance, we detect He~II but not He~I, and we see N~V but not N~IV.  We have applied for
{\it HST} time to obtain the UV data needed to better determine these values, as these will provide additional 
ionization states (for instance, N~IV $\lambda 1718$), and key diagnostics of the stellar wind (e.g.,
C~IV $\lambda 1550$).  The full modeling will be discussed once those data are obtained, or, if we are not
successful in securing UV data, once we have additional optical data.   But the preliminary modeling does show that the observed spectra {\it can} be produced by a single object, and (if our effective temperatures are correct) that the bolometric luminosities, and hence the progenitor masses, would be
normal rather than small.  Why the mass-loss rates are low, and how these stars evolved, remain unanswered questions for the present.  Are they the hitherto unrecognized products of single star evolution, or are binary models needed to produce such objects?  If the latter, then where is the spectroscopic signature of the companion?

\begin{figure}
\plotone{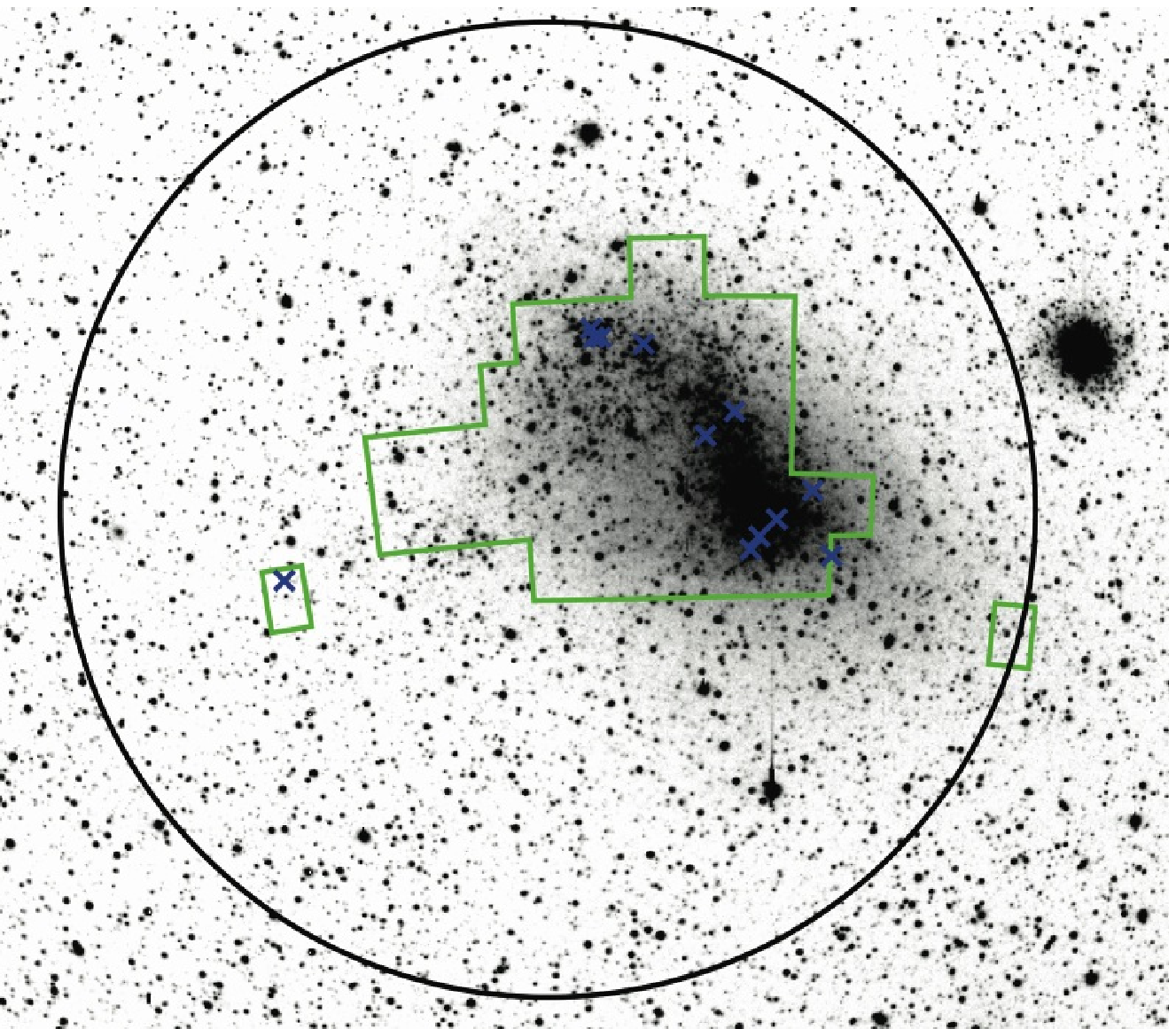}
\caption{\label{fig:SMC}  The WR survey area for the SMC.  The large black circle has a radius of 3\fdg0 and denotes the area of the SMC we plan to survey for WRs.  The blue $\times$'s denote the locations of the known WRs, and the green areas enclose the regions of the 51 $14\farcm8\times22\farcm8$ fields we observed during the first observing run reported here.  The image is the {\it R}-band exposure of the SMC with the ``parking lot" camera described by Bothun \& Thompson (1988).}
\end{figure}

\begin{figure}
\plotone{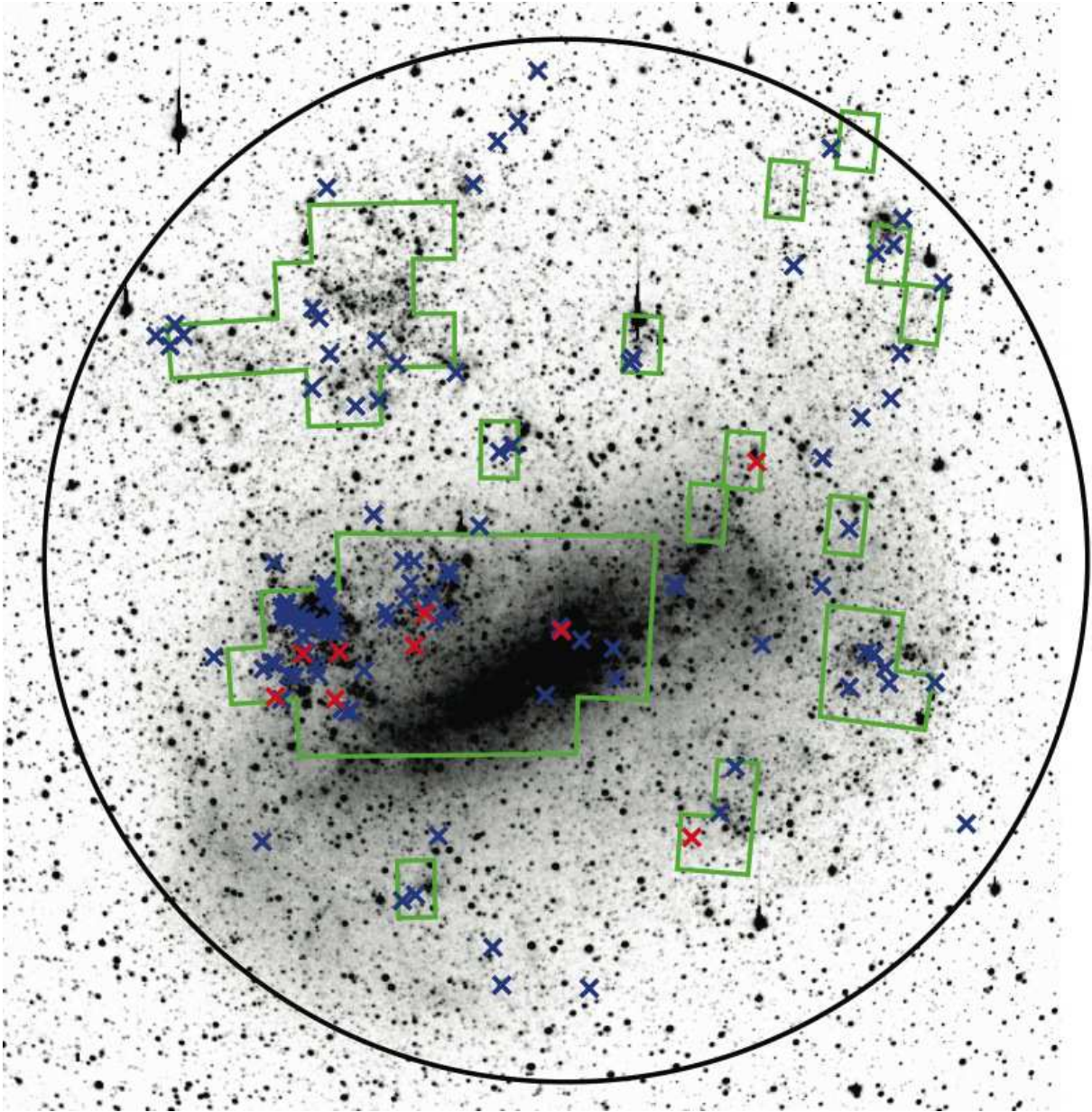}
\caption{\label{fig:LMC}  The WR survey area for the LMC.  The large black circle has a radius of 3\fdg5 and denotes the area of the LMC we plan to survey for WRs.  The blue $\times$'s denote the locations of the known WRs, and the green areas enclose the regions of the 76 $14\farcm8\times22\farcm8$ fields we observed during our first observing run, with red $\times$'s denoting the newly found WRs reported here.  The image is the {\it R}-band exposure of the LMC with the ``parking lot" camera described by Bothun \& Thompson (1988).}
\end{figure}

\begin{figure}
\plotone{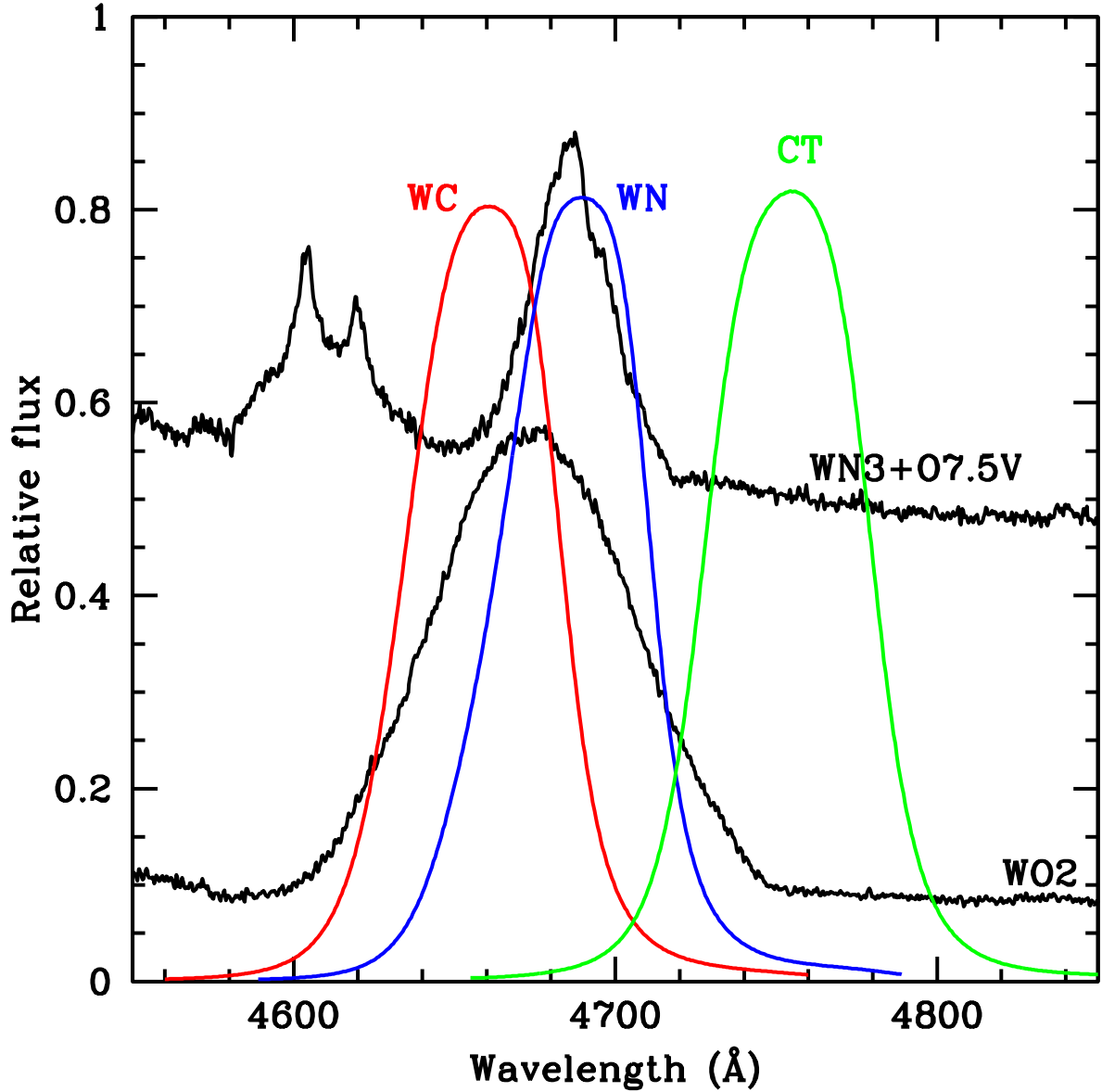}
\caption{\label{fig:test}  Filter bandpasses.  The transmission bandpasses for the {\it WC}, {\it WN}, and {\it CT} filters are shown superimposed on the fluxed spectra of a WN3+O7V pair (above) and a WO2 star (below).  (The WO2 spectrum is very similar to that of a WC4; the only difference is enhanced O VI $\lambda \lambda 3811,34$ outside the region shown.) The WRs are newly discovered to this paper.}
\end{figure}

\begin{figure}
\plotone{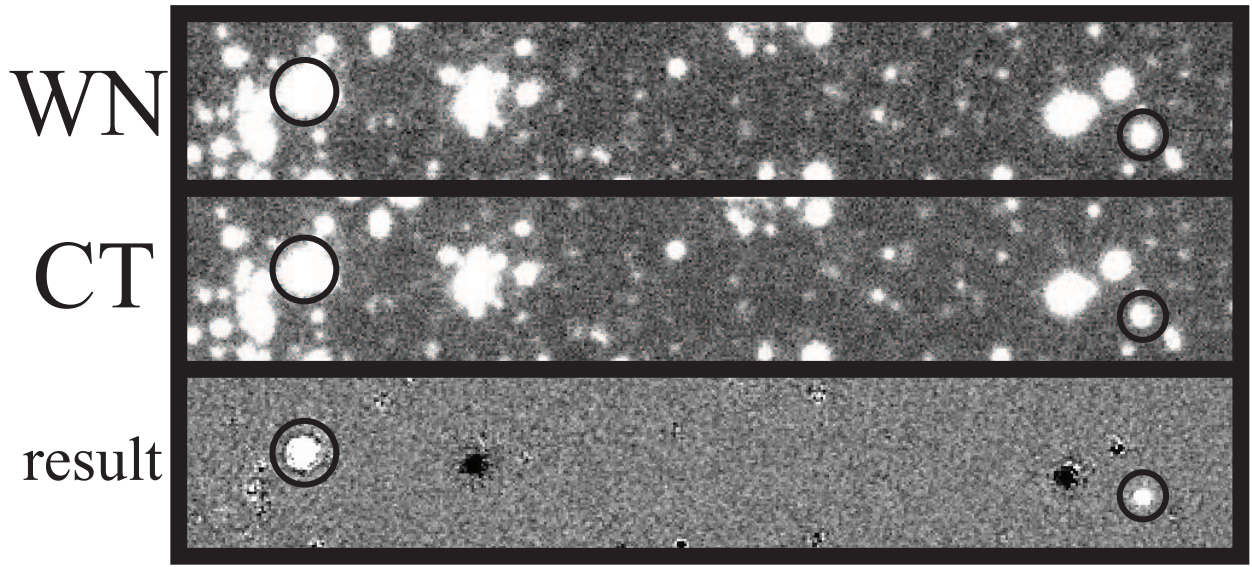}
\caption{\label{fig:wow}  Image subtraction example.   We show a small strip, about 3\farcm0$\times$ 0\farcm5, taken through the {\it WN} filter (upper) and the continuum ({\it CT}) filter (middle), along with the result of image subtraction (bottom).
The two known WRs (circled) stand out in the bottom image as they are brighter in the {\it WN} than in the {\it CT}.
The two black blobs are the residuals left by saturated stars.}
\end{figure}

\begin{figure}
\epsscale{0.4}
\plotone{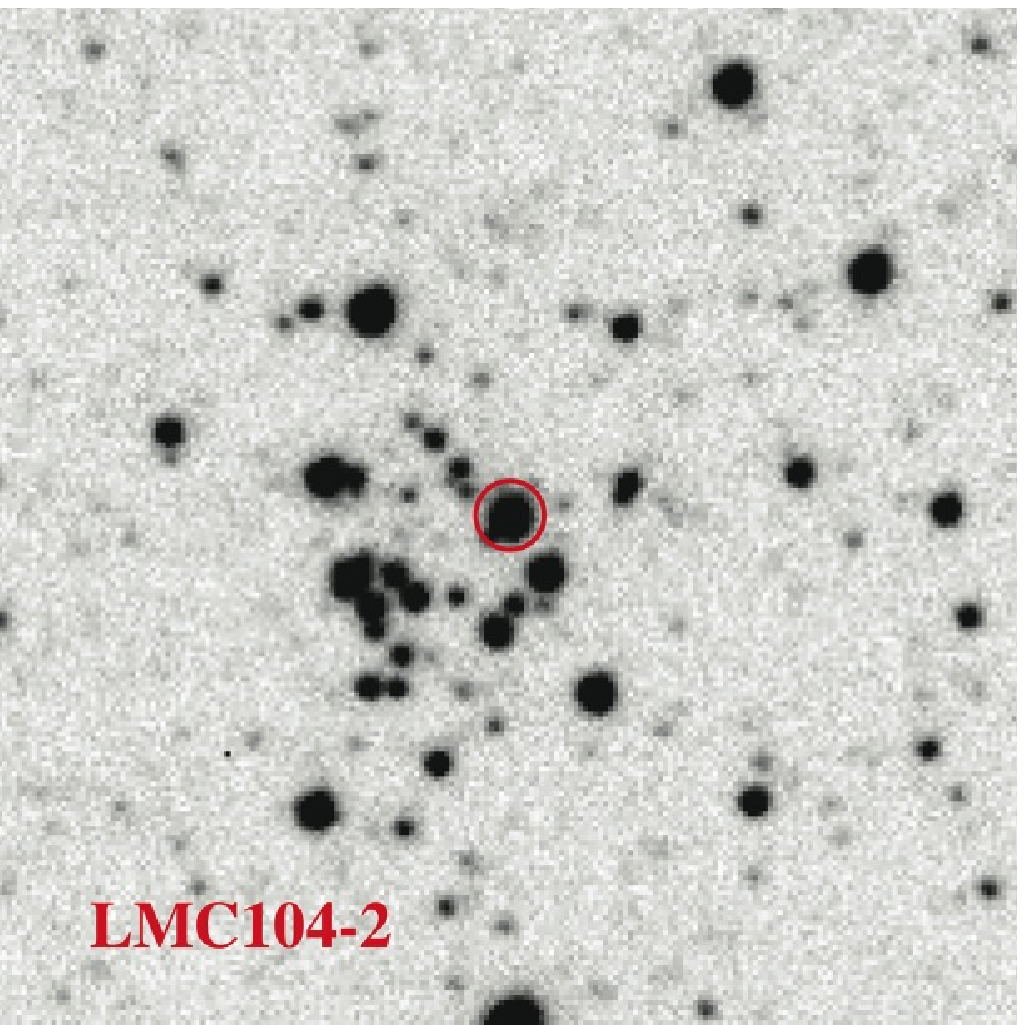}
\plotone{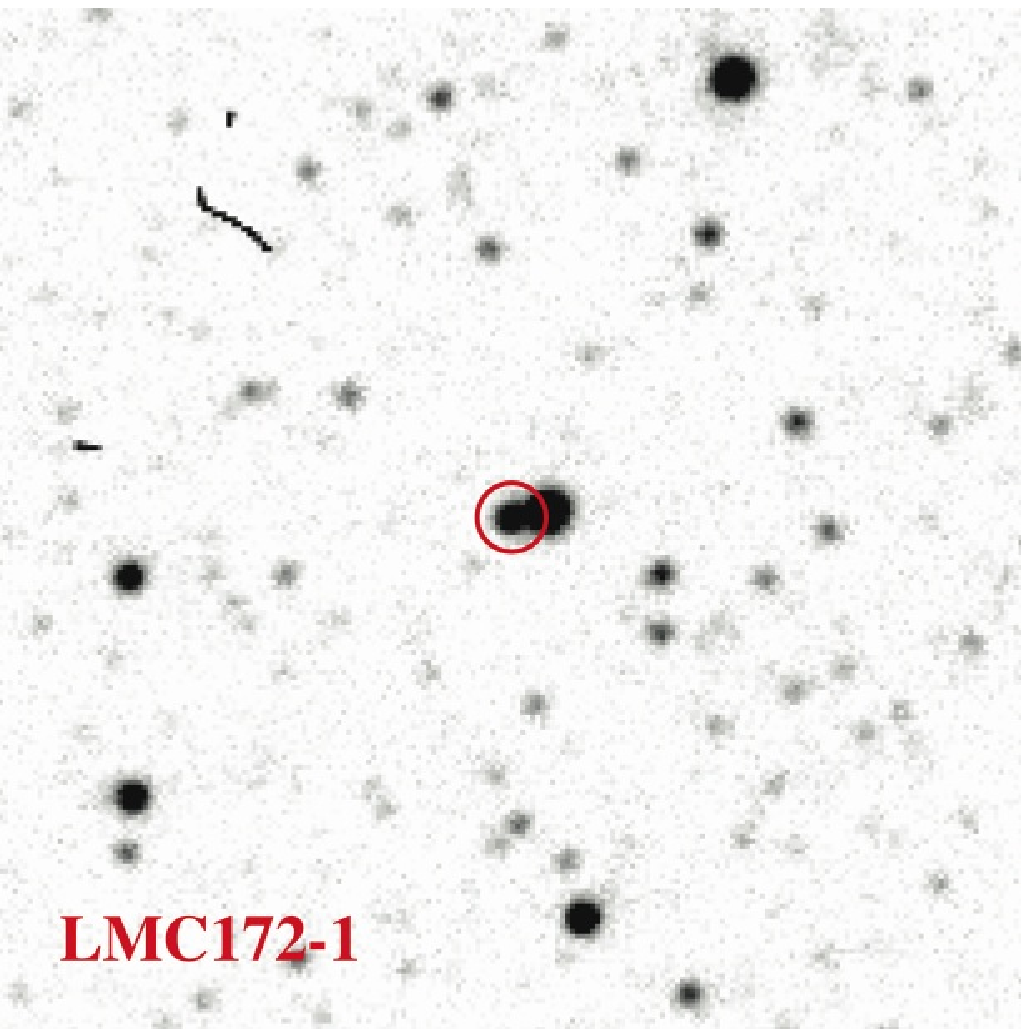}
\plotone{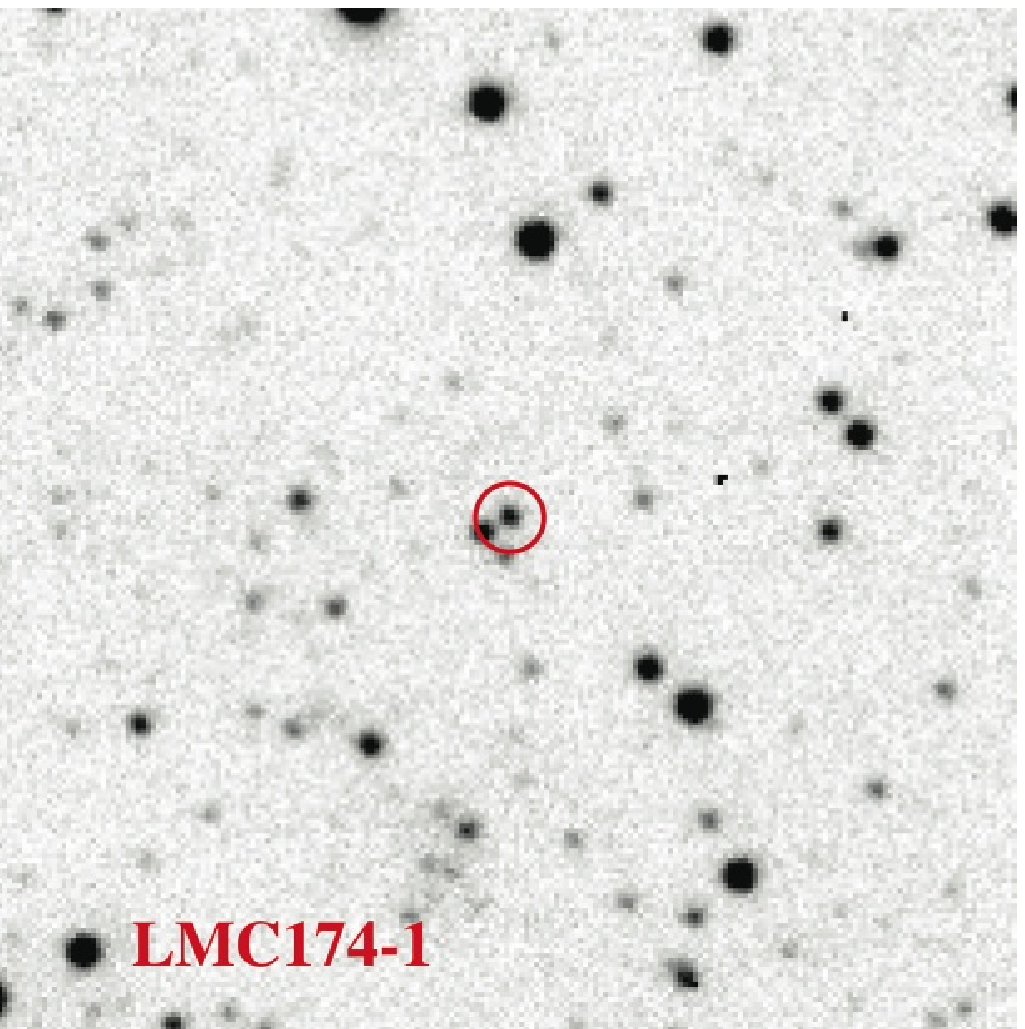}
\plotone{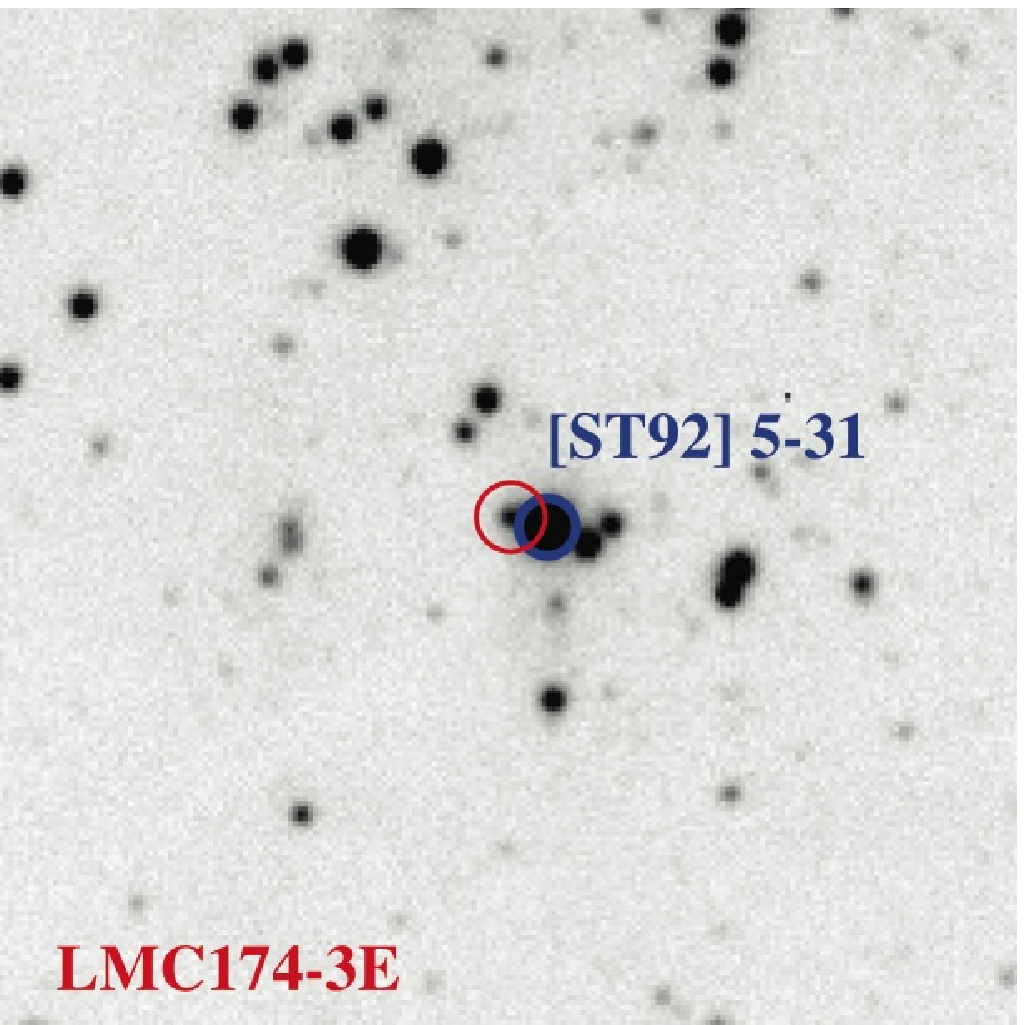}
\plotone{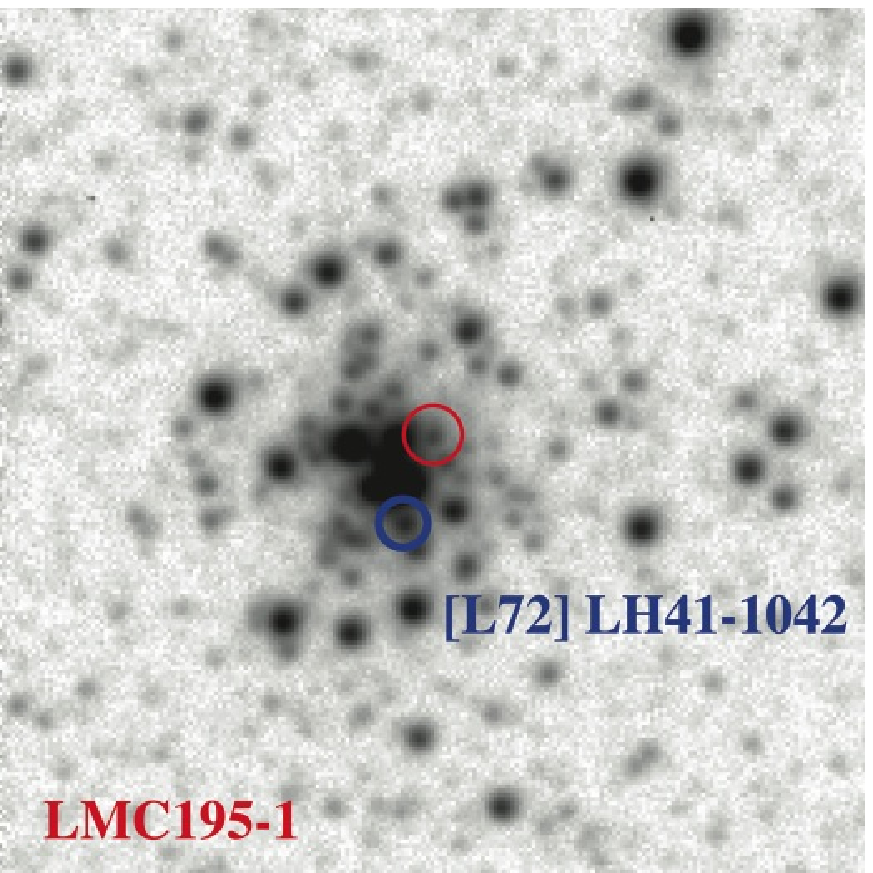}
\epsscale{1.0}
\caption{\label{fig:FCs}  Finding charts for the crowded WRs and Of-type stars in Tables~\ref{tab:WRs} and \ref{tab:Ofs}. The image for LMC195-1 is log scaled; the others are scaled linearly.  The fields are all 1\farcm5 on a side, have N up and E to the left.  The circles are 3\arcsec\ in diameter.}
\end{figure}

\begin{figure}
\epsscale{0.5}
\plotone{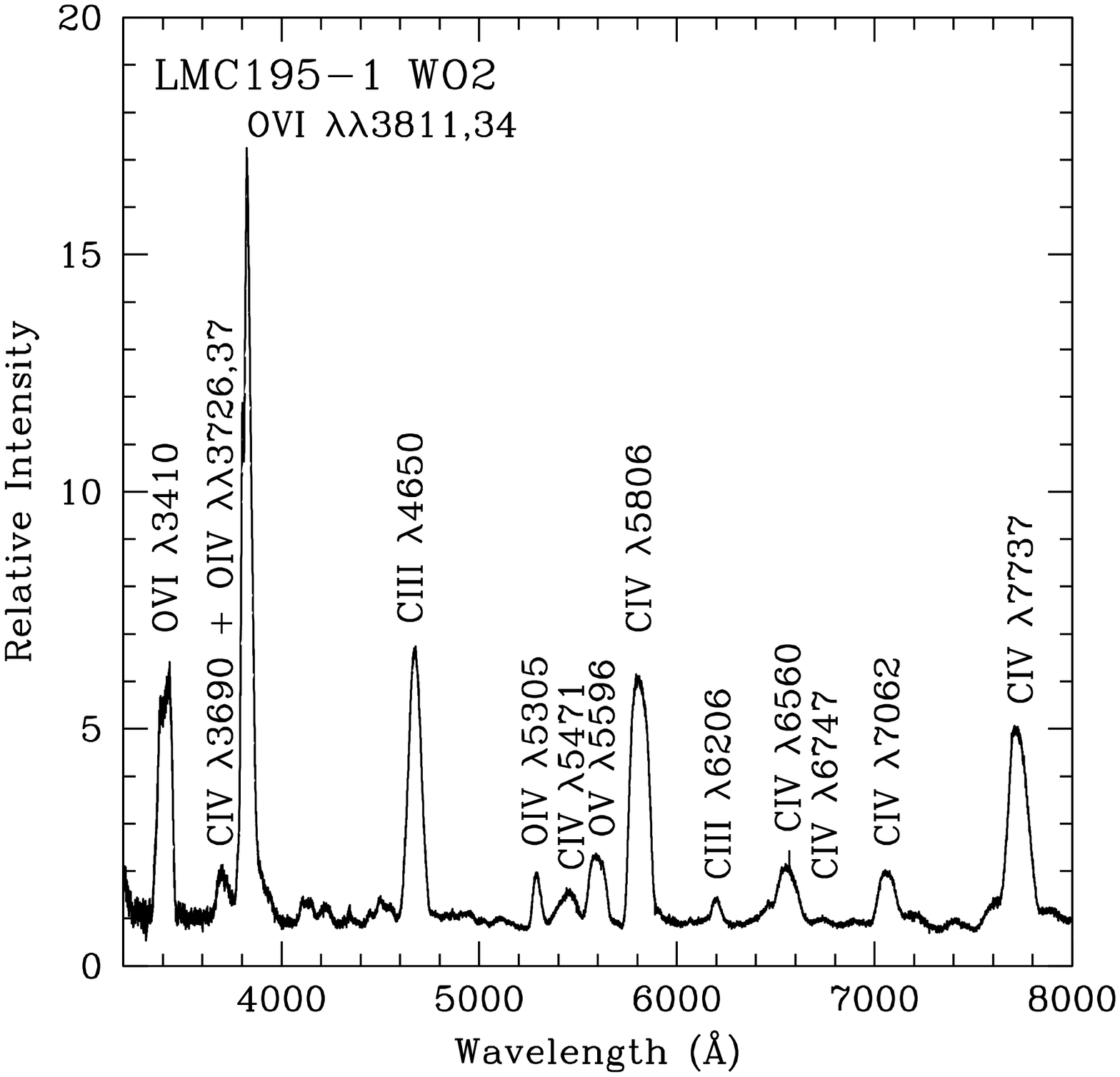}
\plotone{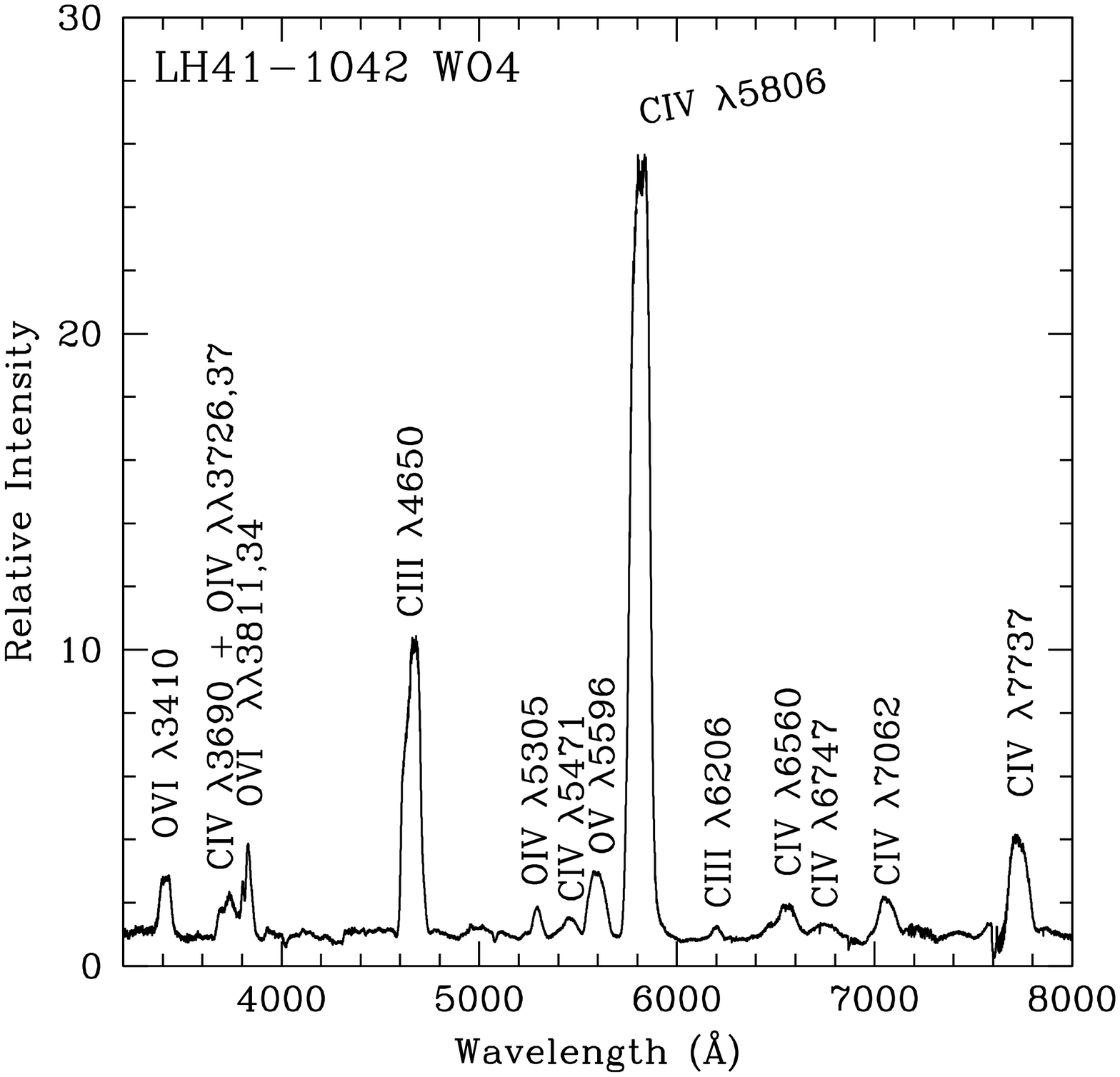}
\epsscale{1.0}
\caption{\label{fig:LMC1951}  Spectra of WO Stars.  {\it Upper:} The principal lines are identified in our spectrum of LMC195-1, our newly found WO2 star.  {\it Lower:} For comparison, we show the spectrum of LH54-1042, a WO4 star (Neugent et al.\ 2012b).
}
\end{figure}

\begin{figure}
\plotone{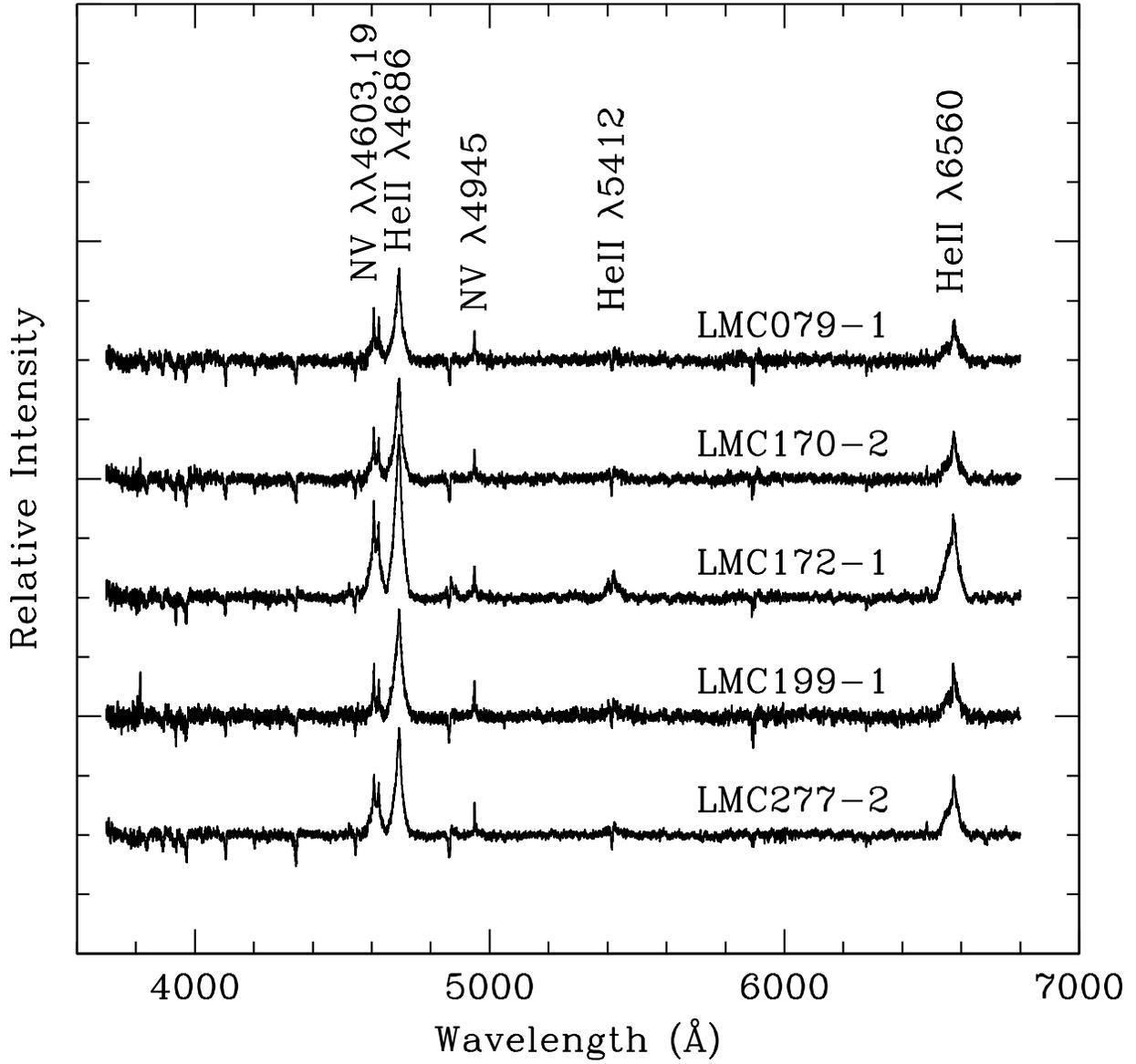}
\caption{\label{fig:O3s}  Spectra of five WN3+O3V stars.  The prominent lines in the yellow and red are marked; detailed identification for the lines in the blue are given in Fig.~\ref{fig:O3}.}
\end{figure}

\begin{figure}
\plotone{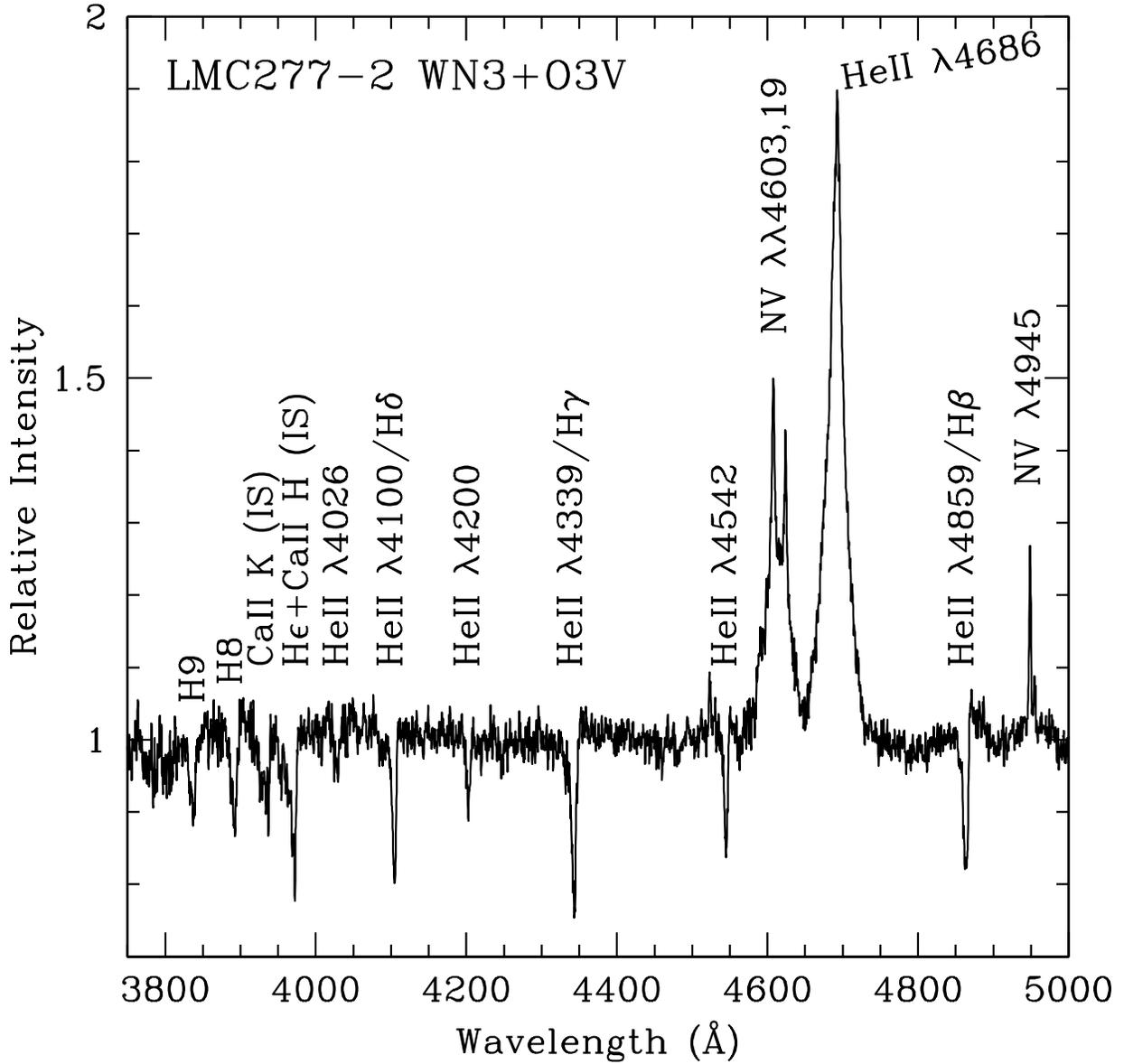}
\caption{\label{fig:O3}  Spectrum of LMC277-2.    The principal spectral features are labeled. The spectrum shown here has been box-car smoothed by 3 pixels for display purposes; the original spectrum has a signal-to-noise of 135 per 1~\AA\ spectral resolution element in the region around 4400-4500~\AA; despite this, there is no sign of He~I $\lambda 4471$.}
\end{figure}

\begin{figure}
\plotone{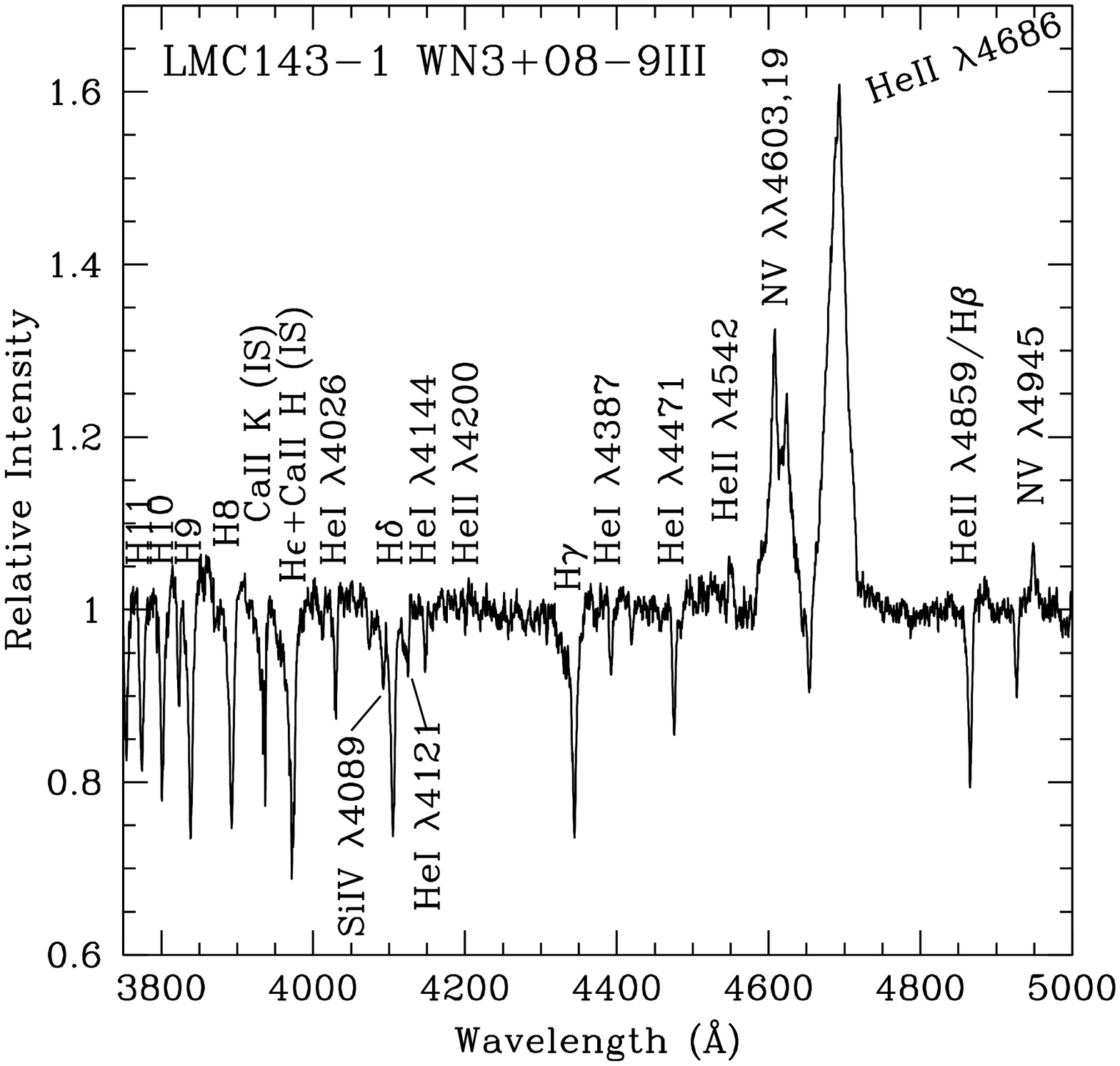}
\caption{\label{fig:LMC1431}  Spectrum of LMC143-1.    The principal spectral features are labeled. The spectrum shown here has been box-car smoothed by 3 pixels for display purposes.}
\end{figure}

\begin{figure}
\plotone{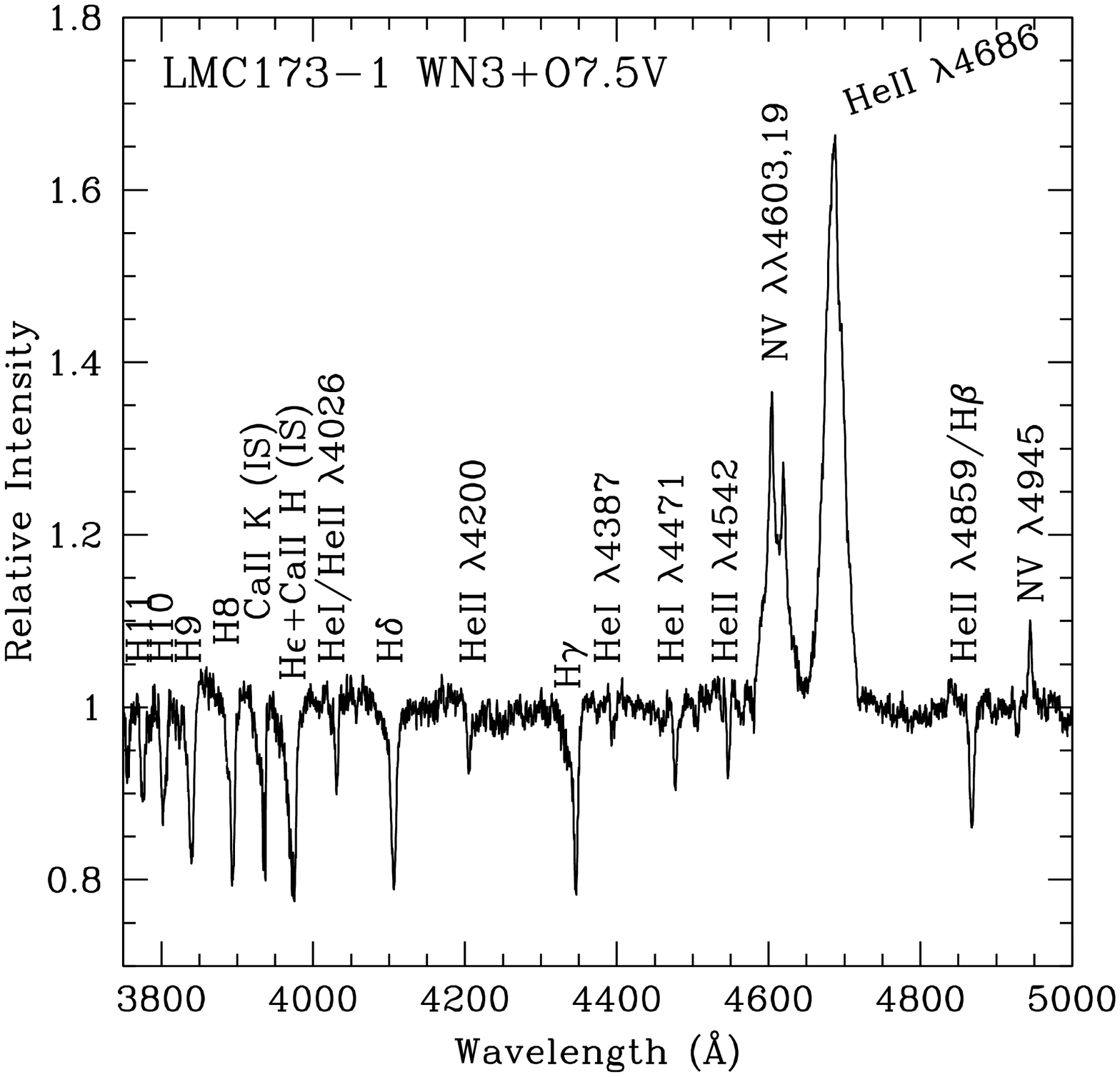}
\caption{\label{fig:LMC1731}  Spectrum of LMC173-1.    The principal spectral features are labeled. The spectrum shown here has been box-car smoothed by 3 pixels for display purposes.}
\end{figure}

\begin{figure}
\plotone{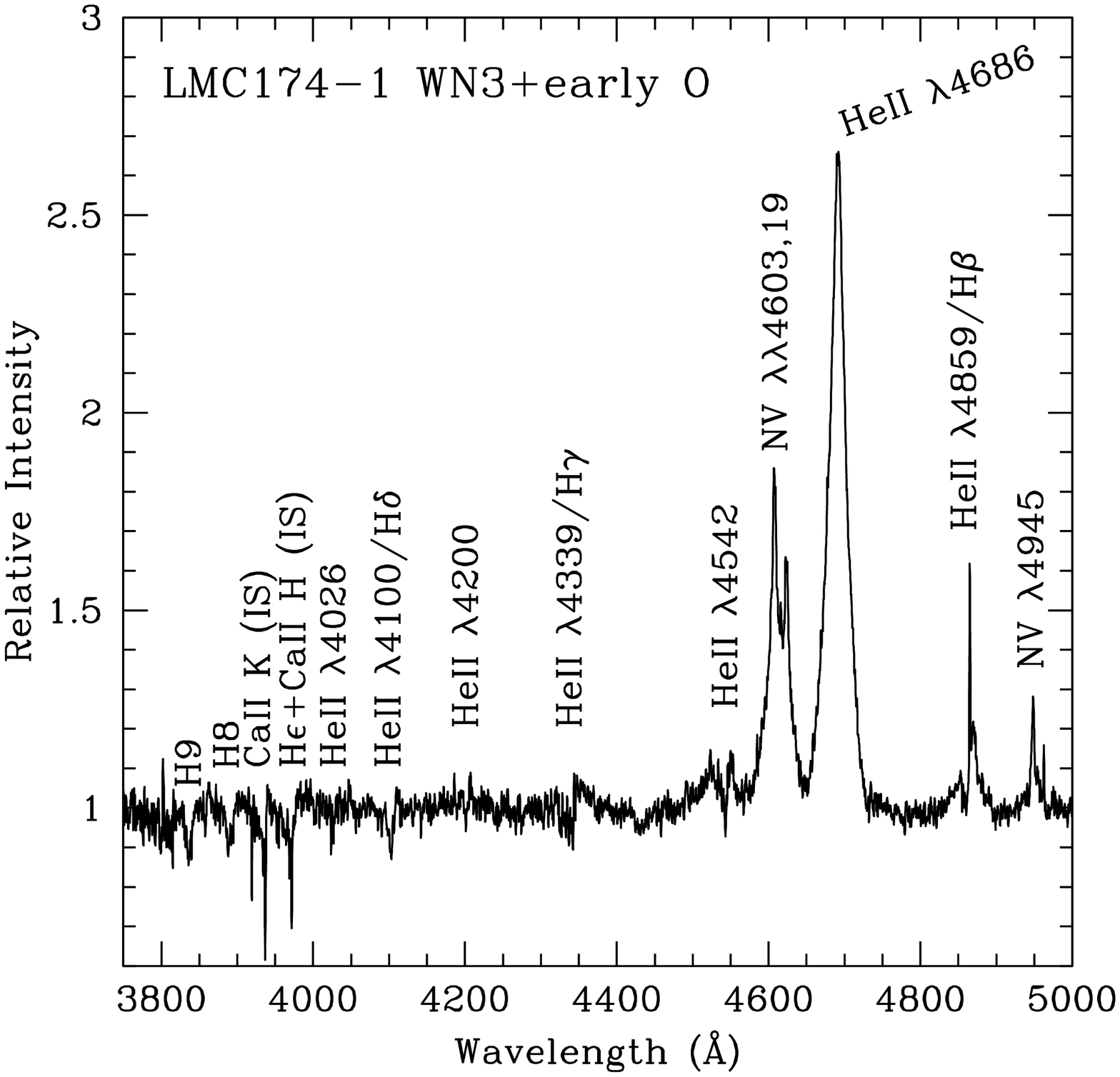}
\caption{\label{fig:LMC1741}  Spectrum of LMC174-1.    The principal spectral features are labeled. The spectrum shown here has been box-car smoothed by 3 pixels for display purposes.}
\end{figure}

\begin{figure}
\epsscale{0.5}
\plotone{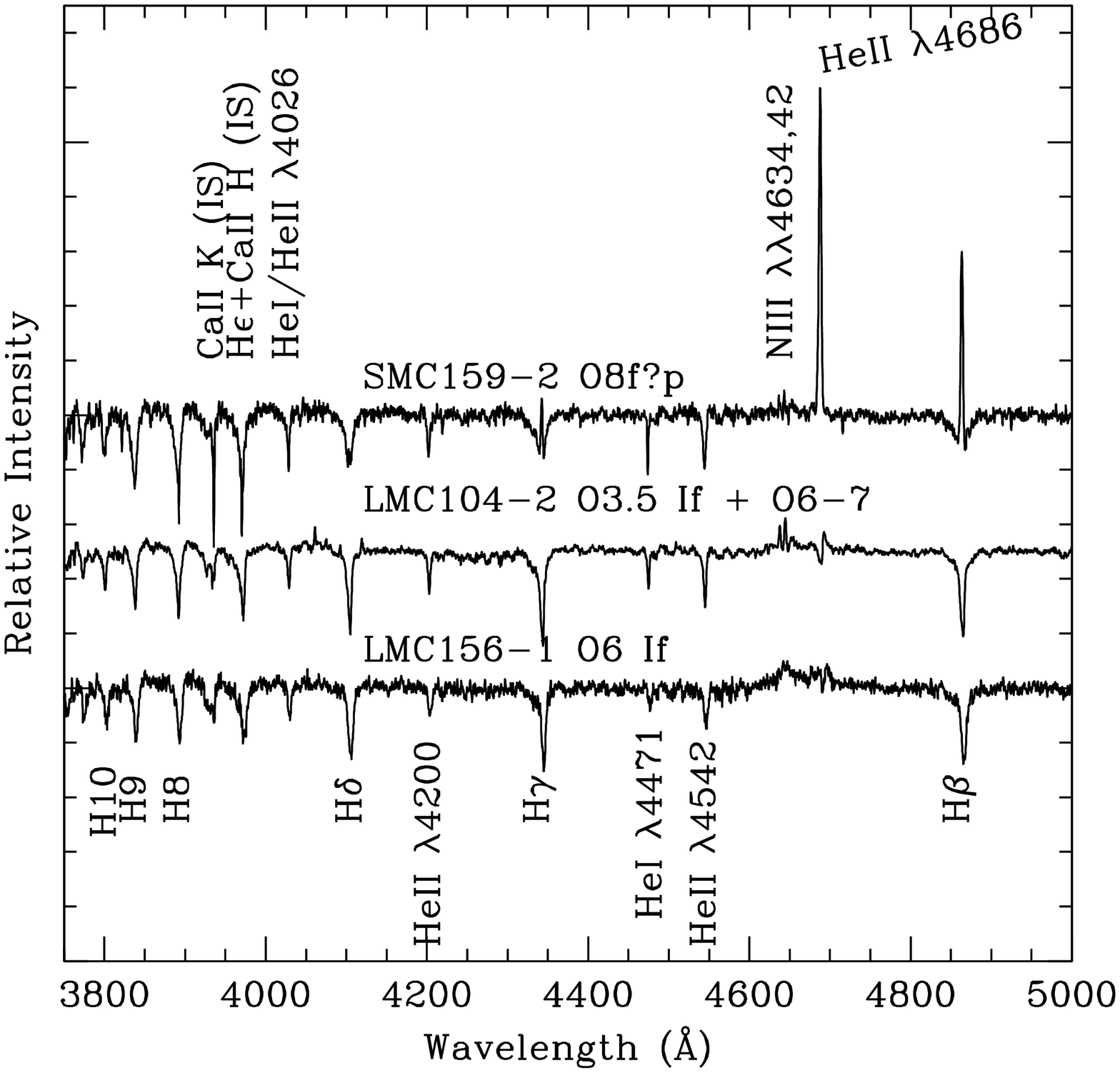}
\plotone{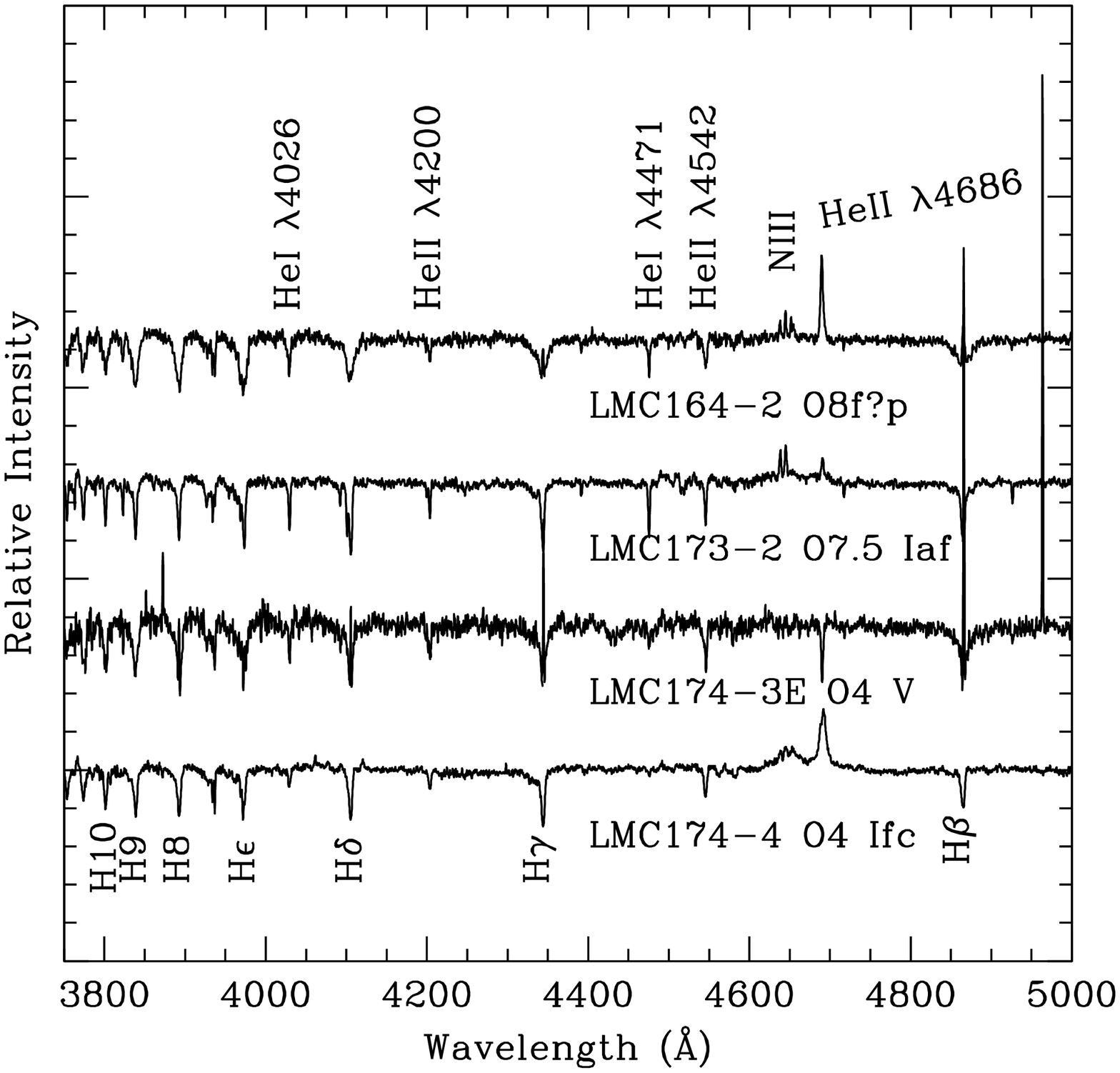}
\epsscale{1.0}
\caption{\label{fig:Ofs}  Spectra of Newly Found O stars.  The principal spectral features are labeled. The spectra shown here have been box-car smoothed by 3 pixels for display purposes.}
\end{figure}

\begin{figure}
\plotone{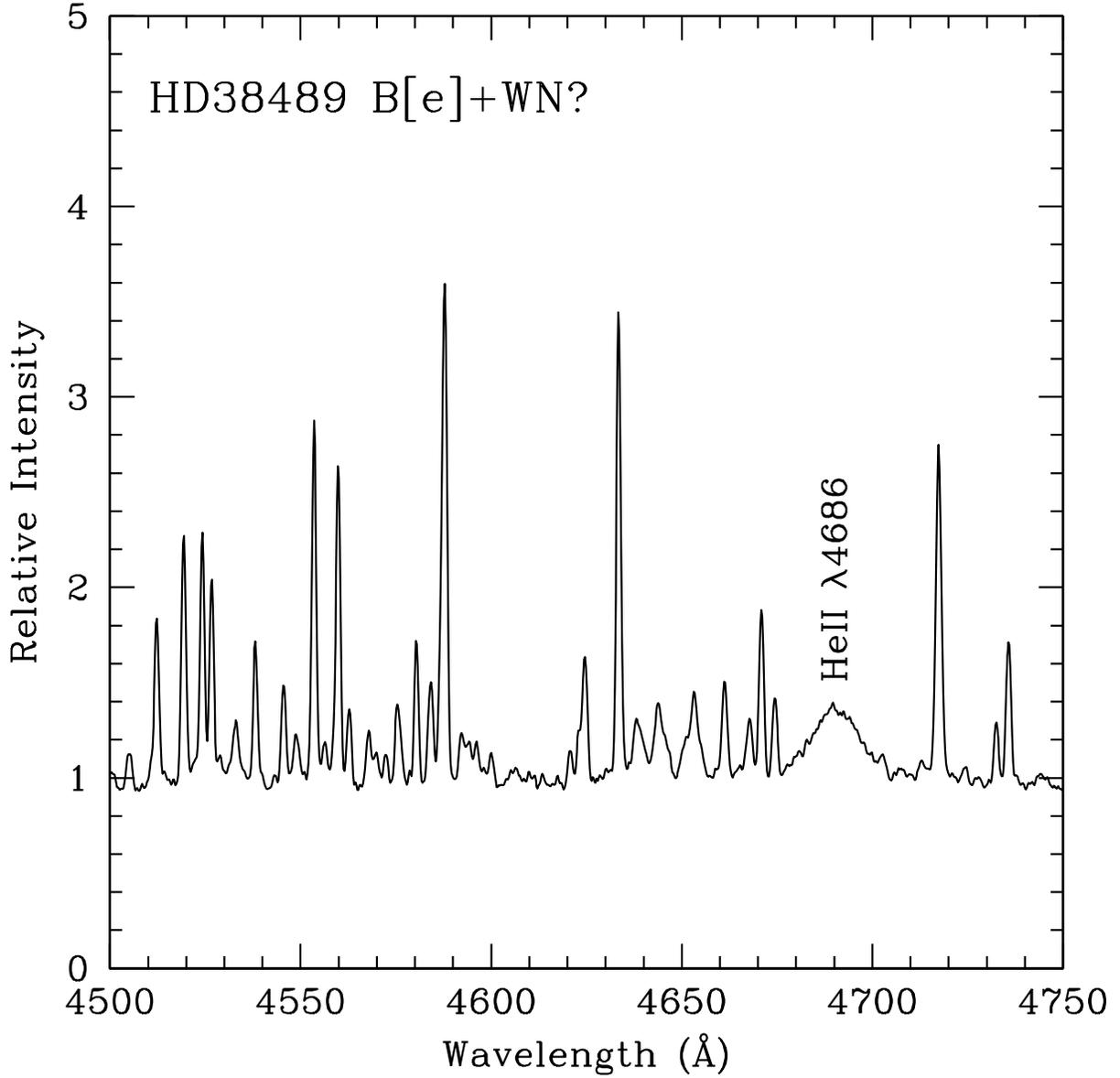}
\caption{\label{fig:HD38489}  Spectrum of LMC 174-5 (HD 38489).    A small portion of our MagE spectrum is shown for this very interesting object. The star shows the classical B[e] narrow emissions (He~I, Fe~II, [Fe~II]) but a very broad He~II $\lambda 4686$ feature.  Compare with Figure 29 of Stahl et al.\ (1985). }
\end{figure}

\begin{figure}
\epsscale{0.42}
\plotone{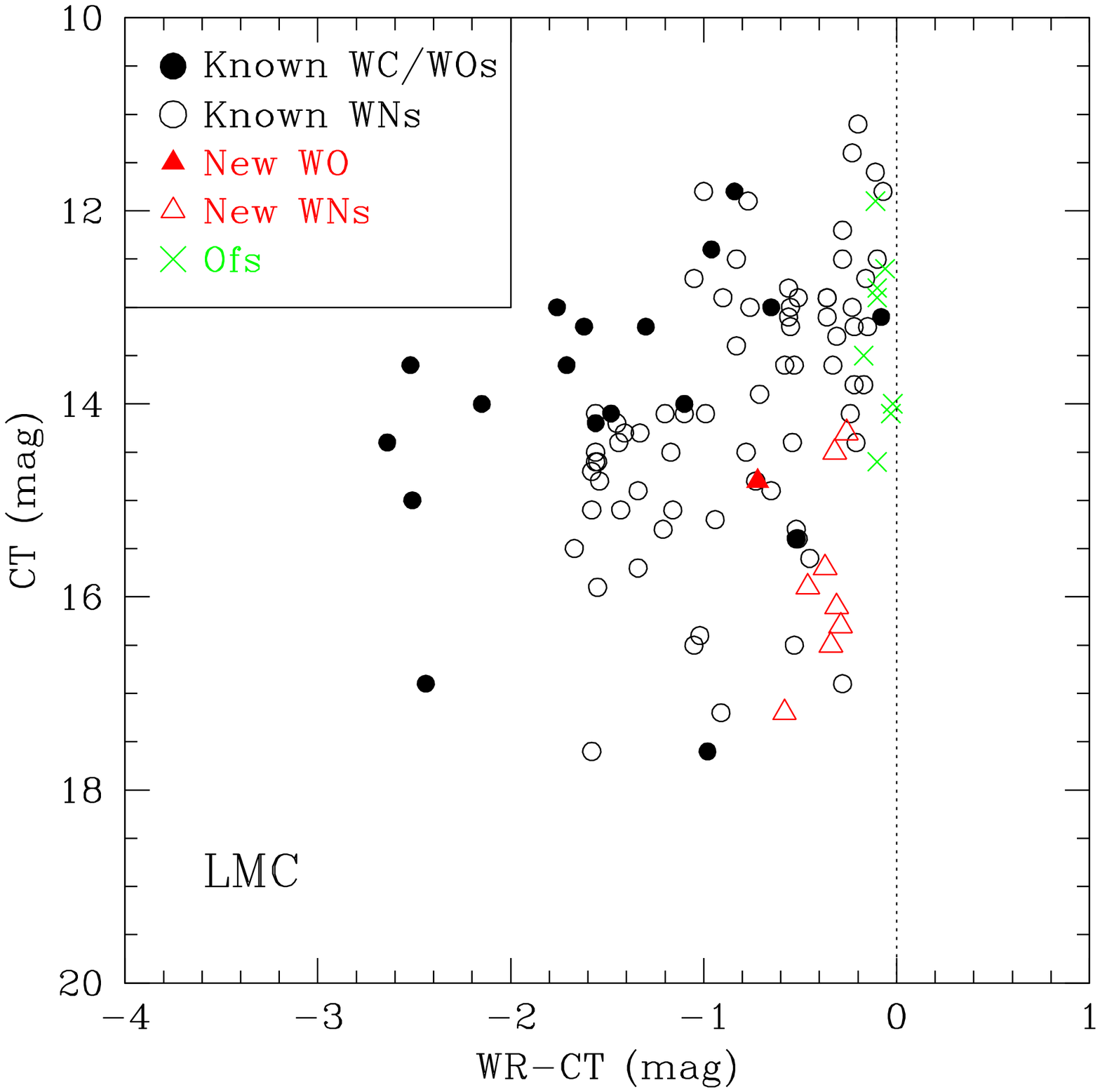}
\plotone{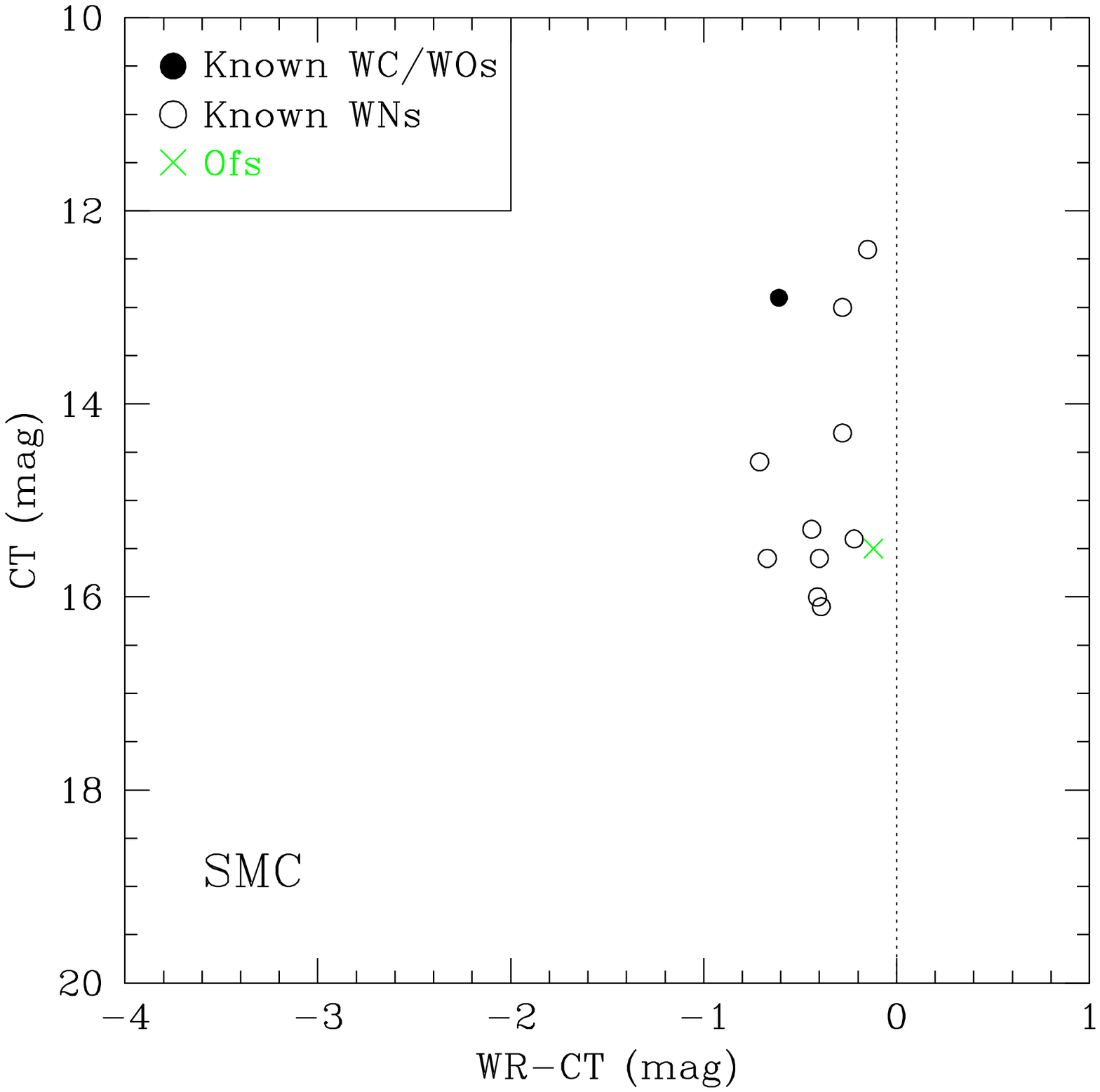}
\plotone{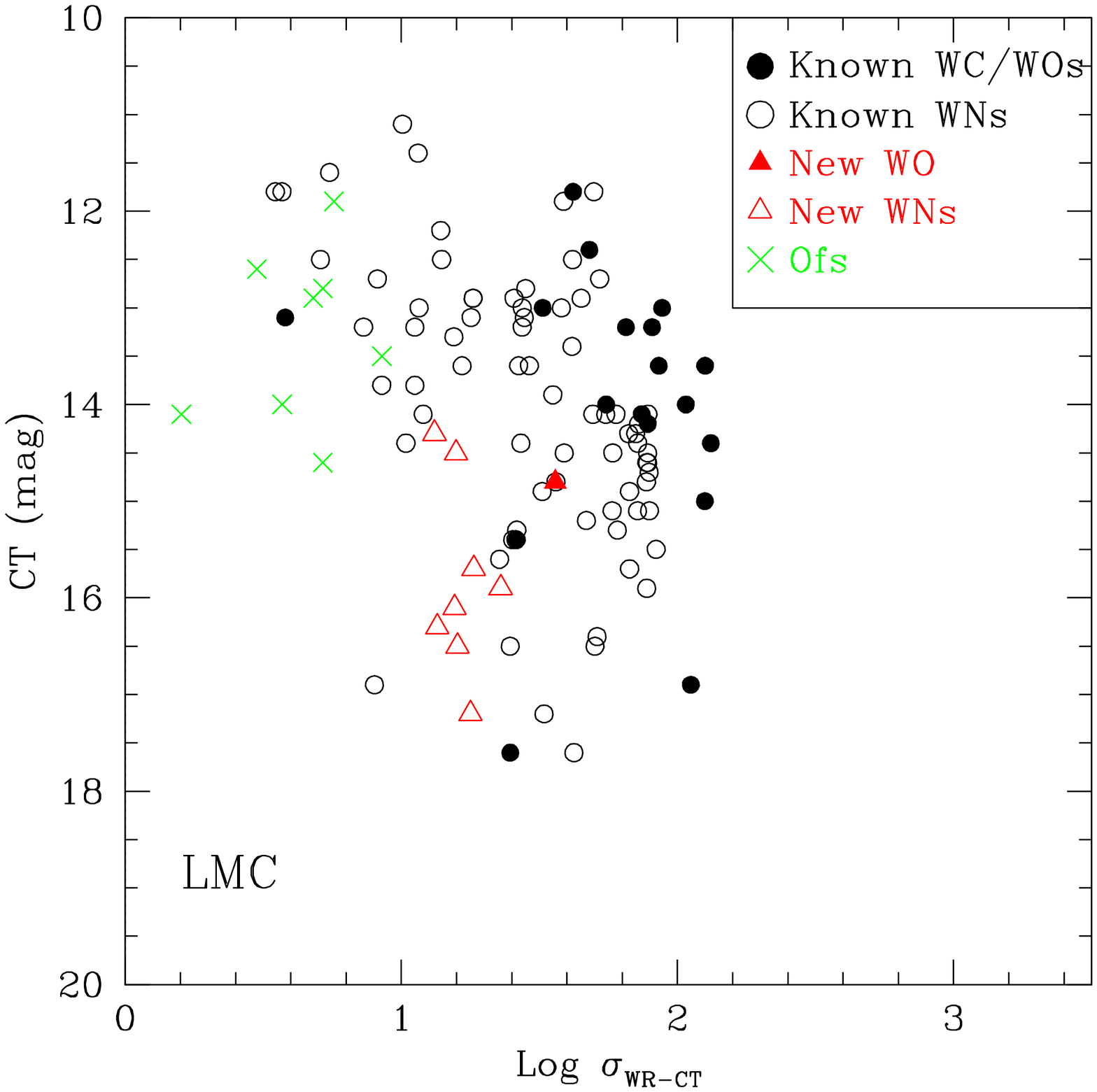}
\plotone{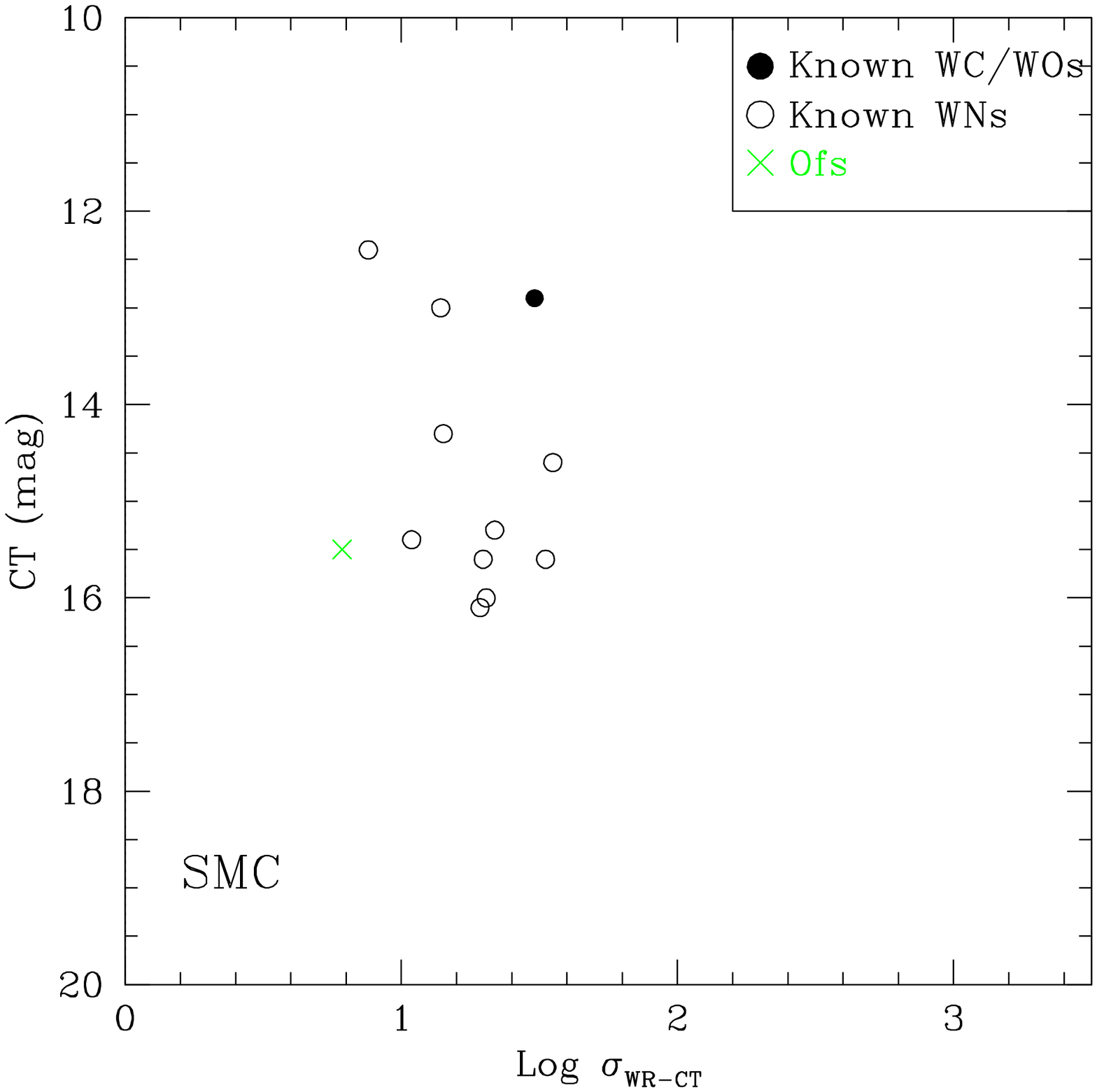}
\epsscale{1.0}
\caption{\label{fig:completeness} Detectability of WRs.  In the upper two figures we plot the continuum magnitude {\it CT} (very similar to $V$) against the magnitude difference between our on-band filter and the continuum filter, {\it WC}$-${\it CT} for the WC and WO stars, and {\it WN}$-${\it CT} for the WN stars. (We refer to this as {\it WR}$-${\it CT}.) In the lower two figures we plot {\it CT} against the logarithm of $\sigma_{WR-CT}$,  where $\sigma_{WR-CT}$ is the significance of the magnitude difference (i.e., the magnitude difference divided by the photometric error). The previously known WCs are shown as filled black circles, and the previously known WNs are shown as open black circles.   The newly found WO star is shown as a filled red triangle, and the newly found WNs are shown as open red triangles.  The Of-stars we know about in our survey fields are shown as green $\times$'s.
}
\end{figure}

\begin{figure}
\epsscale{0.8}
\plotone{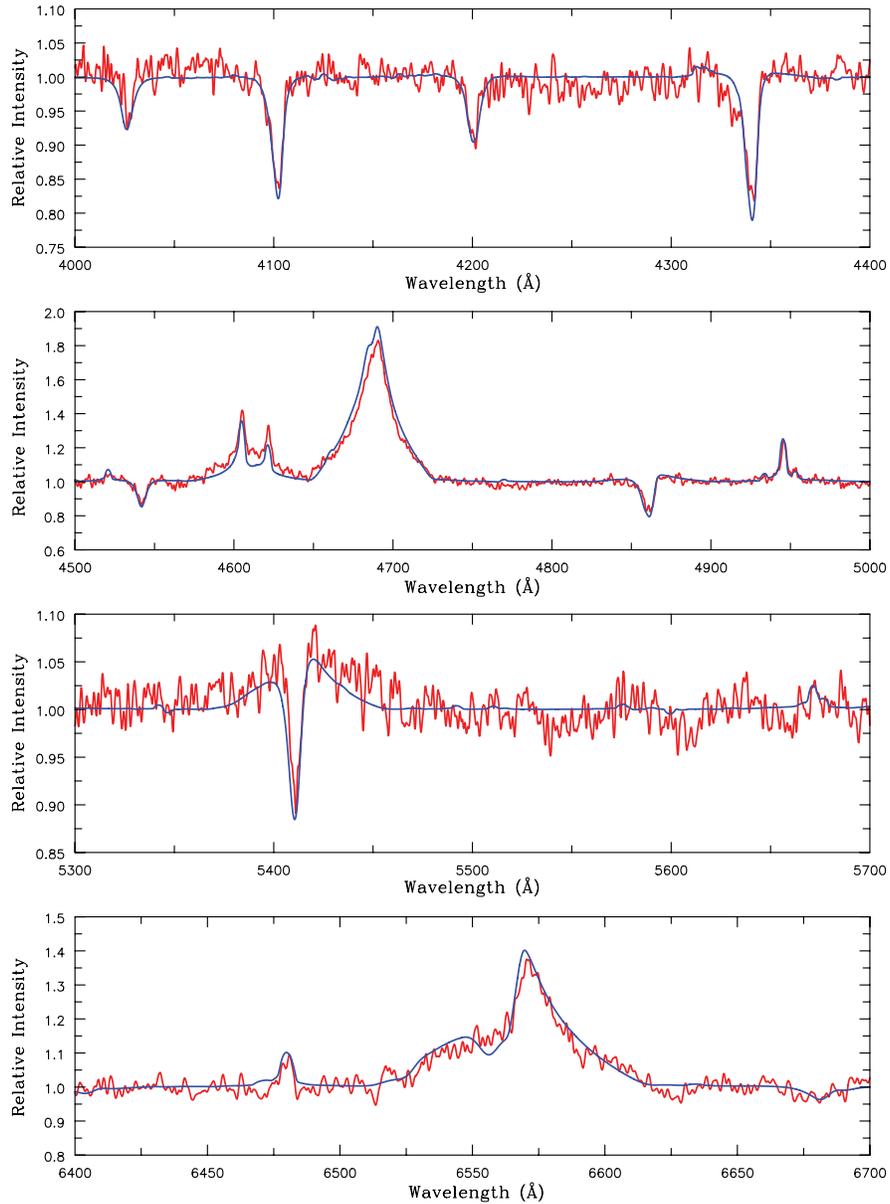}
\epsscale{1.0}
\caption{\label{fig:john} CMFGEN model fit to LMC170-2 .  The synthetic spectrum from the model is shown in blue, and the observed spectrum in red.  The model has an effective temperature of 100,000~K, a He/H number ratio of 1.0, a nitrogen mass fraction of 0.011 (corresponding to 10$\times$ solar using the Anders \&  Grevesse 1989 solar values), and reduced carbon and oxygen abundances (0.05$\times$ solar).
The mass loss rate is $1.2\times 10^{-6} M_\odot$ yr$^{-1}$ with a clumping volume filling factor of 0.1.  For the wind, we used
a terminal velocity of   2,400 km s$^{-1}$, and a standard $\beta=0.8$ wind law parameter.  The bolometric luminosity corresponds to $\log L/L_\odot$ of 5.6, while a $\log g$ of $\sim 5.0$ was adopted.}

\end{figure}

The results from our first observing season have certainly justified in our minds the effort involved in our survey.  As Figs.~\ref{fig:SMC} and \ref{fig:LMC} show, we have so far concentrated on where many WRs were already known.  So it is possible that that we will have a lower success rate next year in terms of finding new ones.  On the other hand, we will not know until we look.  Will any new ones be as equally intriguing as the ones we found this year?  We look forward to more surprises.

\acknowledgements

We thank Nolan Walborn for his useful, constructive referee report, and in particular in suggesting several refinements to our classification of the O-type stars.   Deidre Hunter and an earlier, anonymous referee also helped us
improve the presentation of the material.  We also thank Carlos Contreras for his suggestion to try the image subtraction code {\it HOTPANTS} as well as his help in getting it to execute successfully.  {\it HOTPANTS} was written by Andrew
Becker (University of Washington), who kindly makes the code freely available.   We also thank
Dustin Lang (Carnegie Mellon University) for advice and support for the ``astrometry.net" software.  We are grateful to Francesco Di Mille for obtaining several spectra for us during engineering time on the Clay.
This paper makes use of data products from the Two Micron All Sky Survey, which is a joint project of the University of Massachusetts and the Infrared Processing and Analysis Center/California Institute of Technology, funded by the National Aeronautics and Space Administration and the National Science Foundation.
We would additionally like to thank Brian Skiff for his literature search on potential WR candidates and, as always, the excellent support staff at Las Campanas for all of their help during our observing runs.  This work was supported by the National Science Foundation under AST-1008020 and by Lowell Observatory's research support fund, thanks to generous donations by Mr.\ Michael Beckage and Mr.\ Donald Trantow.  D.J.H. acknowledges support from STScI theory grant HST-AR-12640.01.

{\it Facilities:} \facility{Magellan:Clay(MagE spectrograph)}, \facility{Swope (SITe No. 3 imaging CCD)}

\clearpage
\thispagestyle{empty}
\begin{deluxetable}{l c c c c c c c c c l l }
\rotate
\tabletypesize{\scriptsize}
\tablecaption{\label{tab:WRs} Newly Found WRs}
\tablewidth{0pt}
\tablehead{
\colhead{ID\tablenotemark{a}}
& \colhead{$\alpha_{\rm 2000}$} 
& \colhead{$\delta_{\rm 2000}$} 
& \multicolumn{2}{c}{Other ID}
& \colhead{V\tablenotemark{d}}
& \colhead{{\it CT}} 
& \colhead{$M_V$\tablenotemark{e}} 
& \multicolumn{2}{c}{C~III or He~II} 
& \colhead{Sp.\ Type}  
& \colhead{Comments} \\ \cline{4-5} \cline{9-10}
&&&\colhead{[M2002]\tablenotemark{b}} 
& \colhead{IRSF\tablenotemark{c}}
& & & & \colhead{$\log$(-EW)} 
& \colhead{FWHM} \\
}
\startdata   
LMC079-1 & 05 07 13.33 & $-70$ 33 33.9 & 71747  &  05071333-7033339& 16.06   & 16.3 & -2.8 & 1.3 & 30 &  WN3 + O3 V     & 2 obs \\
LMC143-1 & 05 35 28.52 & $-69$ 40 08.9 & \nodata&  05352853-6940090& 14.09   & 14.1 & -4.8 & 1.2 & 28 &  WN3 + O8-9 III & 2 obs \\
LMC170-2 & 05 29 18.18 & $-69$ 19 43.2 & 143741 &  05291819-6919432& 16.04   & 16.1 & -2.9 & 1.3 & 29 &  WN3 + O3 V     & 2 obs \\
LMC172-1 & 05 35 00.90 & $-69$ 21 20.2 & \nodata&  05350090-6921202& 15.95   & 15.9 & -3.0 & 1.6 & 40 &  WN3 + O3 V     & 2 obs \\
LMC173-1 & 05 37 47.62 & $-69$ 21 13.6 & 169271 &  05374762-6921136& 14.56   & 14.5 & -4.3 & 1.3 & 29 &  WN3 + O7.5 V   & 2 obs \\
LMC174-1 & 05 40 03.57 & $-69$ 37 53.1 & \nodata&  05400357-6937531& 17.11   & 17.2 & -3.0\tablenotemark{f} & 1.7 & 30 &  WN3 + early O  & \\
LMC195-1 & 05 18 10.33 & $-69$ 13 02.5 & \nodata&  05181033-6913025& \nodata & 14.8 & -4.1\tablenotemark{g} & 2.6 & 71 &  WO4            & Crowded (LH41, NGC1910)\\
LMC199-1 & 05 28 27.12 & $-69$ 06 36.2 & \nodata&  05282712-6906362& 16.65   & 16.5 & -2.3 & 0.2 & 8  &  WN3 + O3 V     & \\
LMC277-2 & 05 04 32.65 & $-68$ 00 59.7 & \nodata & 05043264-6800594& 15.83   & 15.7 & -3.1 & 1.4 & 29 &  WN3 + O3 V     & \\
\enddata
\tablenotetext{a}{Designation from the current survey; see Sect.~\ref{Sec-results} for further details.} 
\tablenotetext{b}{Designation from Massey 2002.}
\tablenotetext{c}{Designation from Kato et al.\ 2007.}
\tablenotetext{d}{$V$ from Massey 2002 if cross ID to ``[M2002]"; otherwise, from Zaritsky et al.\ 2004.}
\tablenotetext{e}{Assumes an apparent distance modulus of 18.9 to the LMC; i.e., a distance of 50 kpc and
an average reddening of $A_V=0.4$ except as noted; see Massey et al.\ 2005 Table 1 and references therein.}
\tablenotetext{f}{Adopted a value of $A_V=1.6$ based upon the fluxed spectrum.}
\tablenotetext{g}{Computed based on the {\it CT} magnitude.}
\end{deluxetable}

\clearpage
\thispagestyle{empty}
\begin{deluxetable}{l c c c c c c c c c l l }
\rotate
\tabletypesize{\scriptsize}
\tablecaption{\label{tab:Ofs} Other Interesting Stars}
\tablewidth{0pt}
\tablehead{
\colhead{ID\tablenotemark{a}}
& \colhead{$\alpha_{\rm 2000}$} 
& \colhead{$\delta_{\rm 2000}$} 
& \multicolumn{2}{c}{Other ID}
& \colhead{V\tablenotemark{d}}
& \colhead{{\it CT}} 
& \colhead{$M_V$\tablenotemark{e}} 
& \multicolumn{2}{c}{C~III or He~II} 
& \colhead{Sp.\ Type}  
& \colhead{Comments} \\ \cline{4-5} \cline{9-10}
&&&\colhead{[M2002]\tablenotemark{b}} 
& \colhead{IRSF\tablenotemark{c}}
& & & & \colhead{$\log$(-EW)} 
& \colhead{FWHM} \\
}
\startdata                        
SMC159-2 & 00 49 58.72  &$-73$ 19 28.4 & 15094 &  00495871-7319284 & 15.12 & 15.5 & -4.0 &  0.6 & 4 & O8f?p & \\
LMC104-2 & 05 03 23.28 & $-70$ 20 06.4 & 59329   & 05032328-7020065 & 13.62   & 14.0 & -5.3 & \nodata & \nodata &  O3.5 If + O6-7  & He~II P Cyg \\
LMC156-1 & 04 51 38.38 &$-69$ 29 54.2 & 11248    &  04513837-6929542 & 13.90  & 14.1 & -5.0 & 0.1: & \nodata  & O6 If \\
LMC164-2 & 05 13 49.88 &$-69$ 23 21.7 & 94588    & 05134985-6923216 & 14.37 & 14.6  & -4.5 & -0.2 & 4 & O8f?p\\
LMC173-2 & 05 36 38.77 & $-69$ 27 59.5 & 166084  & 05363879-6927596 & 12.67   & 12.6 & -6.2 & -0.5 & 3 &  O7.5 Iaf      \\
LMC174-3E & 05 39 12.12 &$-69$ 30  36.3 & \nodata & \nodata & \nodata & \nodata & \nodata & \nodata  & \nodata & O4 V & Crowded; see Fig.~\ref{fig:FCs}\\ 
LMC174-4 & 05 39 42.87 & $-69$ 26 52.8 & 173785  & 05394286-6926528 & 12.99   & 12.9 & -5.9 &  0.5 & 10 &  O4 Ifc          \\
LMC174-5& 05 40 13.33 &$-69$ 22 46.5 & \nodata &05401334-6922465& 11.99 &12.4 &  -6.9& 0.7 & 15 & B[e]+WN? &  HD 38489=LHA 120-S134 \\               
\enddata
\tablenotetext{a}{Designation from the current survey; see Sect.~\ref{Sec-results} for further details.} 
\tablenotetext{b}{Designation from Massey 2002.}
\tablenotetext{c}{Designation from Kato et al.\ 2007.}
\tablenotetext{d}{$V$ from Massey 2002 if cross ID to ``[M2002]"; otherwise, from Massey et al.\ 2000}
\tablenotetext{e}{Assumes  apparent distance moduli of 18.9 to the LMC and 19.1 to the SMC, (i.e., distances
of 50 and 59 kpc) from van den Bergh (2000) and $A_V=0.4$ and 0.3, respectively, from Massey et al.\ 1995.}
\end{deluxetable}

\end{document}